\documentclass[]{siamart1116}

\floatstyle{plaintop}
\restylefloat{table}

\usepackage{amsfonts}
\usepackage{amsmath}
\usepackage{graphicx}
\usepackage{epstopdf}
\usepackage{algorithmic}
\usepackage{bm}
\usepackage[]{subfig}
\usepackage[normalem]{ulem}
\usepackage{xcolor}
\usepackage{multirow}
\usepackage{threeparttable}

\ifpdf
  \DeclareGraphicsExtensions{.eps,.pdf,.png,.jpg}
\else
  \DeclareGraphicsExtensions{.eps}
\fi

\numberwithin{theorem}{section}

\newcommand{\TheTitle}{Data-Driven Failure Prediction in Brittle Materials: A Phase-Field Based Machine Learning Framework} 
\newcommand{\TheAuthors}{E. A. Barros de Moraes, H. Salehi, and M. Zayernouri}

\headers{}{}

\author{
  Eduardo A. Barros de Moraes\thanks{Dual PhD Student, Department of Mechanical Engineering \& Department of Computational Mathematics, Science and Engineering, Michigan State University, 428 S Shaw Ln, East Lansing, MI 48824, USA.}
  \and
  Hadi Salehi\thanks{Postdoctoral Associate, Department of Mechanical Engineering, Michigan State University, 428 S Shaw Ln, East Lansing, MI 48824, USA.}
  \and
  Mohsen Zayernouri \thanks{Assistant Professor, Department of Mechanical Engineering \& Department of Statistics and Probability, Michigan State University, 428 S Shaw Ln, East Lansing, MI 48824, USA, Corresponding Author; \email{zayern@msu.edu}}
}

\title{{\TheTitle}\thanks{This work was supported by MURI/ARO Award (W911NF-15-1-0562), the ARO Young Investigator Program Award (W911NF-19-1-0444), the National Science Foundation Award (DMS-1923201), also partially by the AFOSR Young Investigator Program Award (FA9550-17-1-0150).}}

\usepackage{amsopn}

\newtheorem{remark}{Remark}

\ifpdf
\hypersetup{
	pdftitle={\TheTitle},
	pdfauthor={\TheAuthors}
}
\fi

\AtBeginDocument{\let\latexlabel\label}

\begin{document}
	\maketitle

  \begin{abstract}
		Failure in brittle materials led by the evolution of micro- to macro-cracks under repetitive or increasing loads is often catastrophic with no significant plasticity to advert the onset of fracture. Early failure detection with respective location are utterly important features in any practical application, both of which can be effectively addressed using artificial intelligence. In this paper, we develop a supervised machine learning (ML) framework to predict failure in an isothermal, linear elastic and isotropic phase-field model for damage and fatigue of brittle materials. Time-series data of the phase-field model is extracted from virtual sensing nodes at different locations of the geometry. A pattern recognition scheme is introduced to represent time-series data/sensor nodes responses as a pattern with a corresponding label, integrated with ML algorithms, used for damage classification with identified patterns. We perform an uncertainty analysis by superposing random noise to the time-series data to assess the robustness of the framework with noise-polluted data. Results indicate that the proposed framework is capable of predicting failure with acceptable accuracy even in the presence of high noise levels. The findings demonstrate satisfactory performance of the supervised ML framework, and the applicability of artificial intelligence and ML to a practical engineering problem, i.,e, data-driven failure prediction in brittle materials.
	\end{abstract}
	

	\section{Introduction}
	\label{sec:introduction}
	Predictability is essential to any mathematical model for failure and fracture. From early linear elastic fracture mechanic models from \cite{griffith1921vi}, to failure analysis through damage mechanics by \cite{lemaitre2005engineering}, numerical models have improved in scope and complexity to provide realistic simulations of material failure to meet industry goals of safety, and to reduce component weight and production costs. The accurate simulation of the failure process, from crack initiation to propagation until final failure, in a consistent way, while respecting the physics and developing robust numerical methods, is still a challenging task. 
	
	During the last decade, phase-field models have been successfully established as a powerful tool in the study of damage and fatigue. By modeling sharp interfaces through smooth continuous fields, the dynamics of moving boundaries using phase-fields emerged in diverse physical applications, including fluid separation \cite{cahn_free_1958}, solidification \cite{allen_microscopic_1979}, tumor growth \cite{lima_hybrid_2014}, two-phase complex fluid flow \cite{yue_feng_liu_shen_2004}, and fluid-structure interaction \cite{sun_full_2014}. In failure analysis, crack sharpness is modeled through a smooth phase-field indicating the state of the material among fractured, virgin, and intermediate damaged zones, evolving through Allen-Cahn type equations. Examples ranging from brittle \cite{miehe_phase_2010,miehe_thermodynamically_2010,borden_higher-order_2014}, ductile \cite{ambati_phase-field_2015,ambati_phase-field_2016}, and dynamic fracture \cite{borden_phase-field_2012,hofacker_phase_2013} successfully described phenomenological effects such as crack initiation, branching and coalescence. The inclusion of fatigue effects was initially attempted with Ginsburg-Landau free-energy potentials \cite{amendola_thermomechanics_2016} and fractional derivatives \cite{caputo_damage_2015}. A more general framework for damage and fatigue was later developed in a non-isothermal and thermodynamically consistent approach \cite{boldrini2016non,chiarelli2017comparison,haveroth2018comparison}, followed by the emergence of further phase-field models for fatigue  \cite{carrara2019framework,seiler2019efficient}. Within this myriad of different models, solution uncertainty and parametric sensitivity are still influential \cite{kharazmi2019fractional,kharazmi2019operator}, and the predictability of phase-field models for arbitrary conditions is yet a withstanding effort \cite{deintegrated}. One promising approach to address the predictability of numerical models is to use artificial intelligence (AI), which has been consistently expanding its applicability over the years.
	
	AI and machine learning (ML) have been widely used in different engineering  applications such as structural health monitoring \cite{liu2011structure,  santos2016machine,al2016comparison,silva2016novel, wootton2017structural,saeidpour2018parameterized,salehi2018emerging,salehi2018structural,salehi2018damage,salehi2019algorithmic,salehi2019data} and fatigue crack detection \cite{rovinelli2018using,lim2018data, lim2018online}. ML algorithms, in the context of failure analysis, have been used for numerous applications, including phase-field models of polymer-based dielectrics \cite{shen2019phase}, phase-field models of solidification \cite{yabansu2017extraction}, and crystal plasticity \cite{papanikolaou2019spatial}. Another interesting application of ML is to obtain a data-driven representation of free-energy potentials in the atomic scale and upscale it to a phase-field model, using Integrable Deep Neural Networks \cite{teichert2019machine}. Specifically for brittle failure, ML has been recently used to build surrogate models based on explicit crack representation \cite{hunter2019reduced}, and in failure prediction using a discrete crack representation model for high-fidelity simulations that feed an artificial neural networks (ANN) algorithm \cite{moore_predictive_2018,schwarzer2019learning}. Nonetheless, the noted studies have only shown the applicability of ML in failure analysis. Therefore, ML methods have not yet been explored in the context of phase-field models for damage. It is noted that the use of ML leads to a new paradigm of phase-field modeling, where we establish the basis for novel data-driven frameworks, allowing systematic infusion of statistical information \cite{samiee2019fractional} and corresponding uncertainty propagation from micro-scale models and experiments into continuum macroscopic failure models.
	
	In this work, we develop an ML algorithmic framework for failure detection and classification merging a pattern recognition (PR) scheme and ML algorithms applied to a damage and fatigue phase-field model. We consider an isothermal, linear elastic and isotropic material under the hypothesis of small deformations and brittle fracture. We simulate the phase-field model using Finite Element Method (FEM) and a semi-implicit time-integration scheme to generate time-series data of damage phase-field $\varphi$ and degradation function $g(\varphi) = (1-\varphi)^2$ from virtual sensing nodes positioned at different locations across a test specimen. We introduce a PR scheme as part of the ML framework, in which time-series data from FEM node responses are considered as a pattern with a corresponding label. We define multiple labels for ``no failure", ``onset of failure" and ``failure" of the test specimen based on tensile test load-displacement curve and damage threshold concept. Once the patterns representing different states of the material are identified, the proposed ML framework employs $k$-nearest neighbor ($k$-NN) and ANN algorithms to detect the presence and location of failure using such patterns. In this study, we consider different failure types to further assess the performance of the framework. In addition, by introducing noise to the time-series data, we ascertain the robustness of the proposed framework with noise-polluted data, leading to the effective use in failure analysis under high sensitive/uncertain parameters and operators. The findings from this study will pave a way for the development of novel data-driven failure prediction frameworks, which are able to efficiently establish a link among the classification results (i.e., accuracy) and different phase-field model parameters, thus enabling the computational framework to identify those parameters affecting model's accuracy and updating them to achieve the best performance.   
	
	The paper is organized as follows: in \Cref{sec:model} we present the damage and fatigue phase-field model, which is used to generate time-series data for the ML framework. We introduce the data generation procedure and corresponding label definitions in \Cref{sec:data}. We present the ML framework in \Cref{sec:framework}, where we describe the integration of a pattern recognition scheme with the applied classification algorithms, $k$-NN and ANN. We present and discuss the numerical results in \Cref{sec:results}. We then conclude and summarize the paper in \Cref{sec:conclusions}.
	
	\section{Damage and Fatigue Phase-Field Model}
	\label{sec:model}
	
	\subsection{Governing Equations}
	We consider a isothermal phase-field framework for structural damage and fatigue, modeled by a system of coupled differential equations for the evolution of displacement $\bm{u}$, velocity $\bm{v} = \dot{\bm{u}}$, damage $\varphi$ and fatigue $\mathcal{F}$. The damage phase-field $\varphi$ describes the volumetric fraction of degraded material, and takes $\varphi = 0$ for virgin material, $\varphi = 1$ for fractured material, varying between those states, $0 \leq \varphi \leq 1$, as a damaged material. The evolution equation for the damage field is of Allen-Cahn type since the damage and aging effects are non-conservative and non-decreasing, and is derived along with the equations of motion for $\bm{u}$ and $\bm{v}$ through the principle of virtual power and entropy inequalities with thermodynamic consistency \cite{boldrini2016non}. The fatigue field $\mathcal{F}$ is associated to the presence of micro-cracks, and is treated as an internal variable, whose evolution equation is obtained through constitutive relations that must satisfy the entropy inequality for all admissible processes. The geometry is defined over a spatial domain $\Omega \subset \mathbb{R}^d$, $d = 1,2,3$, at time $t \in (0,T]$.
	
	The final form of the governing equations will be defined by the choice of free-energy potentials related to elasticity, damage and fatigue. We consider a linear elastic isotropic material, where the phase-field free-energy takes the usual gradient form:
	
	\begin{equation}
	\label{Eq:free_energy}
	\Psi (\bm{E}, \varphi, \mathcal{F}) 
	= \frac{1}{2}(1-\varphi)^2 \bm{E}^T \mathcal{C} \bm{E} + 
	g_c \frac{\gamma}{2} |\nabla \varphi|^2  
	+ \mathcal{K} (\varphi, \mathcal{F}) ,
	\end{equation}
	
	\noindent where $\bm{E}=\nabla^S\bm{u}$ is the strain tensor, where $\nabla^S\bm{q} = \text{sym}(\nabla \bm{q})$ represents the symmetric part of the gradient of a given vector field $\bm{q}$. Also, $\mathcal{C}$ is the elasticity tensor written in terms of the Young modulus $E$ and Poisson coefficient $\nu$, $g_c$ is the Griffith energy, $\gamma > 0$ is the phase-field layer width parameter, and $\mathcal{K} (\varphi, \mathcal{F})$ is a function that models the damage evolution due to fatigue effects. The first term in \Cref{Eq:free_energy} represents the degraded elastic response, modeled by the choice of degradation function $g(\varphi) = (1-\varphi)^2$. The final set of governing equations, defined over $\Omega \times (0,T]$, becomes:
	
	\begin{equation}
	\label{Eq:governing}
	\left\{
	\begin{array}{l}
	\dot{\bm{u}} = \bm{v}, \\
	\displaystyle
	\dot{\bm{v}}  
	=  
	\, \mbox{div} \, \left( (1-\varphi)^2\frac{ \mathcal{C} }{\rho} \bm{E} \right) 
	+ \frac{b}{\rho}   \, \mbox{div} \, ( \bm{D} )
	-  \frac{\gamma g_c}{\rho} \, \mbox{div} \, ( \nabla \varphi \otimes \nabla \varphi  )+  \bm{f},\\
	\displaystyle
	\dot{\varphi} 
	= \frac{ \gamma g_c}{\lambda} \Delta \varphi  
	+  \frac{1}{\lambda}  (1- \varphi) \bm{E}^T\mathcal{C} \bm{E}
	- \frac{1}{\lambda \gamma}  [ g_c \mathcal{H}' (\varphi) + \mathcal{F} \mathcal{H}_f' (\varphi) ],
	\\
	\displaystyle
	\dot{\mathcal{F}} = -  \frac{ \hat{F}}{\gamma}   \mathcal{H}_f (\varphi) ,
	\end{array}
	\right.
	\end{equation}
	
	\noindent subjected to appropriate initial and boundary conditions, which depend on the physical problem. Either displacement or stress are known at the boundaries, in addition to considering $\nabla\varphi \cdot \bm{n} = 0\,\, \text{on}\,\, \partial \Omega$. Moreover, the $\otimes$ operator denotes the outer product, the infinitesimal strain rate tensor is represented by $\bm{D}=\nabla^S \bm{v}$, and parameters $b$ and $\rho$ are the material's viscous damping and density, respectively. We construct $\lambda$ such that the rate of change of damage increases with damage (see e.g., \cite{lemaitre2005engineering}):
	
	\begin{equation}
	\label{Eq:lambda}
	\dfrac{1}{\lambda} = \dfrac{c}{(1 + \delta - \varphi)^\varsigma} ,
	\end{equation}
	
	\noindent where $c,\,\varsigma > 0$ are material dependent, and $\delta > 0$ is a small constant to avoid numerical singularity. 
	
	 The potentials $\mathcal{H}(\varphi)$ and $\mathcal{H}_f(\varphi)$ model the damage transition from $0$ to $1$ as fatigue changes from zero to $g_c$. We take their (ordinary) derivatives with respect to $\varphi$ to obtain potentials $\mathcal{H}'(\varphi)$ and $\mathcal{H}_f'(\varphi)$. Further details on fatigue potentials can be found in \cite{boldrini2016non}. Choosing the transition to be continuous and monotonically increasing, suitable choices for the potentials are:
	
	\noindent\begin{minipage}{.55\linewidth}
		\begin{equation}
		\label{Eq:potentials}
		\mathcal{H} (\varphi) =
		\begin{cases}
		0.5 \varphi^2   & \mbox{for} \; 0 \leq \varphi \leq 1 ,
		\vspace{0.1cm} 
		\\
		0.5 + \delta (\varphi -1)  &  \mbox{for} \; \varphi > 1 ,
		\vspace{0.1cm}
		\\
		- \delta \varphi &   \mbox{for} \; \varphi < 0 .
		\end{cases}
		\end{equation} 	
	    \end{minipage}%
        \begin{minipage}{.45\linewidth}
		\begin{equation}
		\label{Eq:potentials2}
		\mathcal{H}_f (\varphi) =
		\begin{cases}
		- \varphi & \mbox{for} \; 0 \leq \varphi \leq 1,
		\vspace{0.1cm} 
		\\
		-1 & \mbox{for} \; \varphi > 1,
		\vspace{0.1cm}
		\\
		\hspace{0.25cm} 0 & \mbox{for} \;  \varphi < 0.
		\end{cases}
		\end{equation}
        \end{minipage}
	
	\noindent The evolution of fatigue $\mathcal{F}$ is controlled by $\hat{F}$, related to the formation and growth of micro-cracks that occur in cyclic loadings. We note that being a measure of energy accumulated in the microsctructure, fatigue variable $\mathcal{F}$ grows even under monotonic loading. The form of $\hat{F}$ depends on the absolute value of the power related to stress in the virgin material:
	
	\begin{equation} 
	\label{Eq:fhat}
	\hat{F} =  a  (1-\varphi) \left|\left(\mathcal{C}\bm{E} + b \bm{D}  \right):\bm{D}\right|,
	\end{equation}
	
	\noindent where the parameter $a$ in this case is chosen to give a linear dependence of the power of stress.

	\subsection{Discretization}
	We discretize \Cref{Eq:governing} in space using linear finite element method (FEM), where the semi-discrete form is obtained through Galerkin method. For detailed derivation of the spatial discretization in 2D, we refer to \cite{deintegrated}. We denote $\ddot{\hat{\bm{u}}} = \dot{\hat{\bm{v}}}$ and write the semi-discrete form for an element $k$ as

    \begin{equation}
    \label{Eq:discrete}
    \left\{
    \begin{array}{l}
    \displaystyle
    \bm{M}^k\,\ddot{\hat{\bm{u}}}^k = \, 
    \bm{K}_{u}^k \, \hat{\bm{u}}^k +
    \bm{K}_{v}^k \, \hat{\bm{v}}^k + 
    \bm{w}_a^k +
    \bm{M}^k\,\hat{\bm{f}}^k ,\\ 
    \bm{M}^k_\varphi\,\dot{\hat{\bm{\varphi}}}^k = \,
    \displaystyle
    {\left( \bm{P}_{\varphi}^k + \bm{K}_c^k \right)} \,\hat{\bm{\varphi}}^k + 
    \bm{w}_b^k + \bm{w}_c^k ,\\
    \displaystyle
    \bm{M}^k_\mathcal{F}\,\dot{\hat{\bm{\mathcal{F}}}}^k = \bm{w}_d^k , \,
    \end{array}
    \right.
    \end{equation}

    \noindent where $\bm{M}$, $\bm{M}_\varphi$ and $\bm{M}_\mathcal{F}$ are mass matrices related to displacement, damage and fatigue, respectively. In the equation of motion, $\bm{K}_{u}$ is the damage-degraded elasticity stiffness matrix, $\bm{K}_{v}$ is the viscous damping matrix and $\bm{w}_a$ is related to gradient of damage. In the damage evolution equation, $\bm{P}_{\varphi}$ includes the Laplacian and potential $\mathcal{H}'(\varphi)$. The influence of displacement in damage is represented by $\bm{K}_c$ and $\bm{w}_b$. The potential $\mathcal{H}_f'(\varphi)$ affects the $\bm{w}_c$ operator, and $\bm{w}_d$ is the operator on the right-hand side of fatigue evolution equation. We obtain the global form of operator matrices by the standard assembly operation, and we drop the superscript $k$ in the global sense.
	
    We further discretize the operators in time, adopting a staggered, semi-implicit time integration scheme, where nonlinear terms are treated explicitly, thus avoiding the use of iterative methods. Let the solution time interval $[0,T]$ be split in discrete time steps $t_n$ with time increments of size $\Delta t = t_{n+1} - t_n > 0$, $n=0, 1, \dots $. For the global approximation of any field variable $\bm{q}$, we denote $\bm{q}_{n+1} =  \hat{\bm{q}}(t_{n+1})$. 
	
	We first solve the damage evolution equation and obtain $\bm{\varphi}_{n+1}$ using the backward Euler scheme, where we treat $\lambda$, displacement, and fatigue explicitly, using values from time step $t_n$. Evolution of damage is then obtained by solving the linear system

    \begin{equation}
    \label{Eq:euler}
    {\left[ {\bm{M}_\varphi}-\Delta t {\left( \bm{P}_{\varphi} + \bm{K}_c \right)} \right]}{\bm{\varphi}}_{n+1} =  {\bm{M}_\varphi} {\bm{\varphi}}_{n} + \Delta t {\left(\bm{w}_b + \bm{w}_c\right)}.
    \end{equation}
    
    After we update the damage field, we use Newmark method to solve displacement and velocity in the equation of motion. We denote acceleration and velocity at time $t_{n+1}$ by 

    \begin{eqnarray}
    && \ddot{{\mathbf{u}}}_{n+1} = \alpha_1 \left( \mathbf{u}_{n+1} - \mathbf{u}_{n} \right) - \alpha_2 \dot{\mathbf{u}}_{n} - \alpha_3 \ddot{\mathbf{u}}_{n} \label{Eq:approx_acceleration}\\
    && \dot{\mathbf{u}}_{n+1} = \alpha_4 \left( \mathbf{u}_{n+1} - \mathbf{u}_{n} \right) + \alpha_5 \dot{\mathbf{u}}_{n} + \alpha_6 \ddot{\mathbf{u}}_{n}, \label{Eq:approx_velocity}
    \end{eqnarray}
    
    \noindent with $\alpha_i$, $i = 1,2,\dots,6$ written in terms of standard Newmark coefficients $\tilde{\gamma}$ and $\tilde{\beta}$:
    
    \begin{equation*}
    \begin{split}
    \refstepcounter{equation} \latexlabel{eq:Eqsalpha1}
    \refstepcounter{equation} \latexlabel{eq:Eqsalpha2}
    \refstepcounter{equation} \latexlabel{eq:Eqsalpha3}
    \refstepcounter{equation} \latexlabel{eq:Eqsalpha4}
    \refstepcounter{equation} \latexlabel{eq:Eqsalpha5}
    \refstepcounter{equation} \latexlabel{eq:Eqsalpha6}
    &\alpha_1 =  \dfrac{1}{\tilde{\beta}{\Delta t}^2}, \,\,
    \alpha_2 =  \dfrac{1}{\tilde{\beta} \Delta t}, \,\,
    \alpha_3 =  \dfrac{1 - 2\tilde{\beta}}{2\tilde{\beta}}, \,\, \\
    &\alpha_4 =  \dfrac{\tilde{\gamma}}{\tilde{\beta}\Delta t}, \,\,
    \alpha_5 =  1 - \dfrac{\tilde{\gamma}}{\tilde{\beta}} \,\,\,
    \textmd{and} \,\,\,
    \alpha_6 =  \left( 1 - \dfrac{\tilde{\gamma}}{2\tilde{\beta}} \right) \Delta t.
    \end{split}
    \tag{\ref{eq:Eqsalpha1}-\ref{eq:Eqsalpha6}}
    \end{equation*}
    
    The discrete form of the equation of motion for computing displacements becomes

    \begin{equation}
    \label{Eq:motion}
    \begin{split}
    \left[ \alpha_1 \bm{M} - \bm{K}_u - \alpha_4 \bm{K}_v \right] \bm{u}_{n+1} &= \bm{M} \left[ \alpha_3 \ddot{\bm{u}}_{n}+ \alpha_2 \dot{\bm{u}}_{n} + \alpha_1 \bm{u}_{n}\right] \\&+ 
    \bm{K}_v \left[ \alpha_6 \ddot{\bm{u}}_{n} + \alpha_5 \dot{\bm{u}}_{n} - \alpha_4 \bm{u}_n\right] + \bm{w}_a + \bm{M} \bm{f}_{n+1}.
    \end{split}
    \end{equation}
    
    Then, we update the acceleration and velocity fields using Equations (\ref{Eq:approx_acceleration}) and (\ref{Eq:approx_velocity}), respectively. When imposing prescribed displacement $\bar{\bm{u}}(t_{n+1})$ we also prescribe corresponding velocity and acceleration at the boundaries using
    
    \begin{equation*}
    \refstepcounter{equation} \latexlabel{eq:Eqsup1}
    \refstepcounter{equation} \latexlabel{eq:Eqsup2}
    \bar{\ddot{\bm{u}}}_{n+1} =  \dfrac{d^2}{dt^2}\bar{\bm{u}}(t_{n+1}) 
    \quad \textmd{and} \quad
    \bar{\dot{\bm{u}}}_{n+1} =  \dfrac{d}{dt}\bar{\bm{u}}(t_{n+1}) 
    \tag{\ref{eq:Eqsup1}-\ref{eq:Eqsup2}},
    \end{equation*}
    \noindent where the bar symbol represents the prescribed degrees of freedom. 
    
    Finally, we use the Trapezoidal method to update the fatigue field obtaining
    
    \begin{equation}
    \label{eq:fatigue_evolution_rule_2}
    {\bm{\mathcal{F}}}_{n+1} = {\bm{\mathcal{F}}}_{n} + \dfrac{\Delta t}{2} \bm{M}^{-1}_\mathcal{F}
    \left[ \bm{w}_d \left( \bm{u}_{n+1},\bm{v}_{n+1},\bm{\varphi}_{n+1} \right) + \bm{w}_d \left( \bm{u}_{n},\bm{v}_{n},\bm{\varphi}_{n} \right) \right].
    \end{equation}
    
\begin{figure}[t]
		\centering
		\subfloat[Geometry and boundary conditions.]{\includegraphics[width=0.49\textwidth]{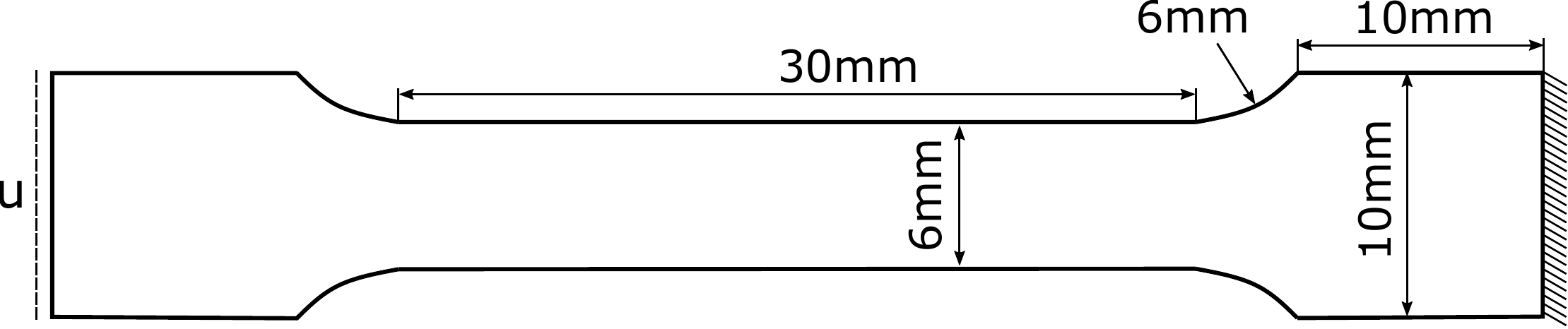}}\hfill
		\subfloat[Mesh and virtual sensor nodes.]{\includegraphics[width=0.49\textwidth]{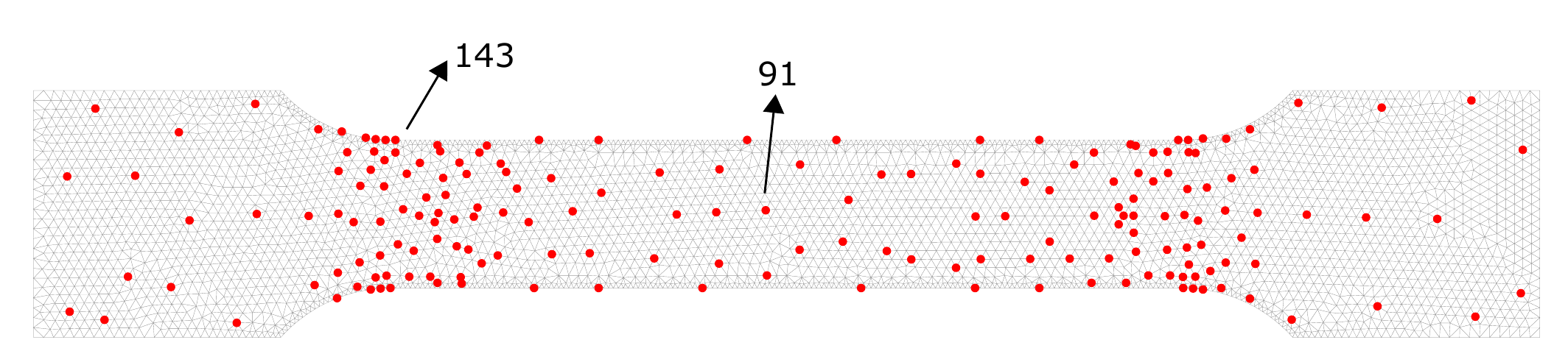}}
		\caption{Description of geometry and boundary conditions for the tensile test specimen, along with finite element mesh and sensor layout for time-series generation. We highlight two sensor nodes that show different time-series behaviors.}
		\label{fig:geo_mesh}
	\end{figure}

	We consider the tensile test specimen without notch depicted in \Cref{fig:geo_mesh}. We discretize it with a finite element mesh consisting of 3912 nodes and 7236 linear triangle elements, with smallest element size of $0.614\ mm$. We constrain one end, and apply a prescribed displacement of $4.5\times 10^{-4}\ m/s$, with time increments of $\Delta t = 5\times 10^{-4}\ s$, at the other end. We study a material with Young modulus $E = 160\ GPa$, Poisson coefficient $\nu = 0.3$, and density $\rho = 7800\ kg/m^3$, under plane stress conditions with thickness of $h = 5\ mm$. The rate of change of fatigue $a$ is $5 \times 10^{-7}\ m^2$, and viscous damping $b$ is $1 \times 10^{8}\ Ns/m^2$. The remaining parameters $\gamma$ (phase-field layer width), $g_c$ (Griffith energy), and $c$ (rate of change of damage), are chosen in order to construct a set of different representative cases. We focus on those parameters to build the cases because they are the most sensitive and give more uncertainty in damage evolution \cite{deintegrated}.

	\section{Data Processing}
	\label{sec:data}
	In this section, we highlight how to obtain time-series data from phase-field simulations to train and test the learning algorithms. Further, we explore different possibilities of label definitions in the context of failure prediction based on the simulation results. 
	
	\subsection{Time-Series Data Generation}
	\label{subsec:time_series}
	
	To generate time-series data, virtual sensing nodes are considered at different locations of the specimen, as shown in \Cref{fig:geo_mesh}. This sensor layout is simply chosen to provide a coarse-to-fine (variable) resolution for the ML framework to calibrate/train and classify time-series data.

\begin{figure}[t]
		\centering
		\subfloat[Case 1, $t = 0.48\ s$.]{\includegraphics[width=0.45\textwidth]{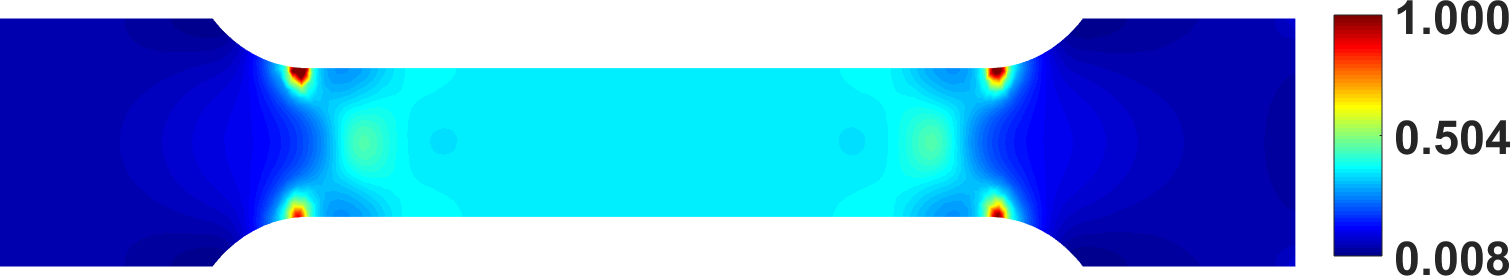}}\hfill
		\subfloat[Case 2, $t = 0.48\ s$.]{\includegraphics[width=0.45\textwidth]{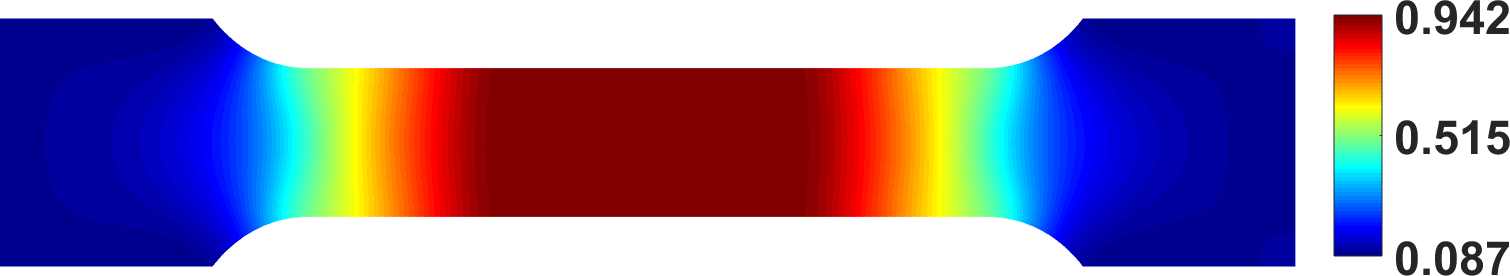}}\\
		\subfloat[Case 3, $t = 0.62\ s$.]{\includegraphics[width=0.45\textwidth]{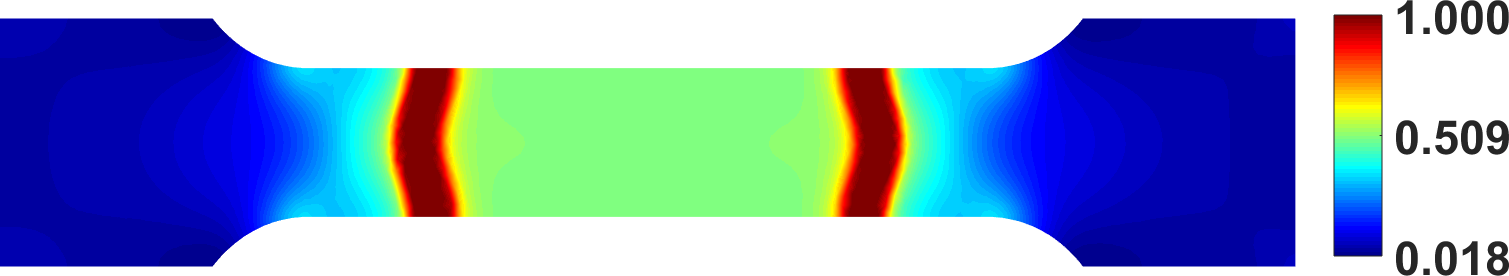}}\hfill
		\subfloat[Case 4, $t = 0.78\ s$.]{\includegraphics[width=0.45\textwidth]{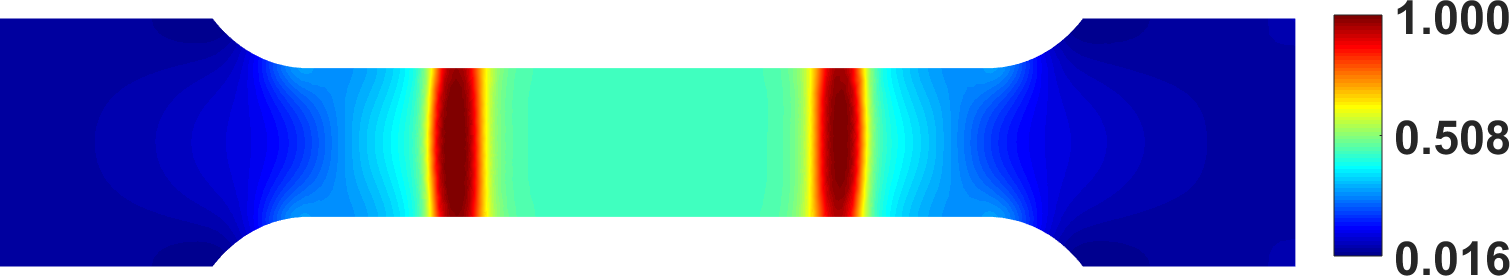}}\\
		\subfloat[Case 5, $t = 0.62\ s$.]{\includegraphics[width=0.45\textwidth]{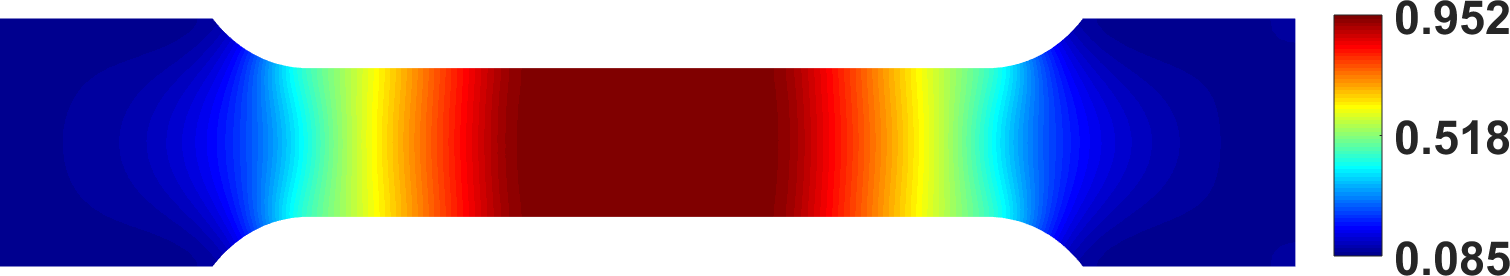}}\hfill
		\subfloat[Case 6, $t = 0.64\ s$.]{\includegraphics[width=0.45\textwidth]{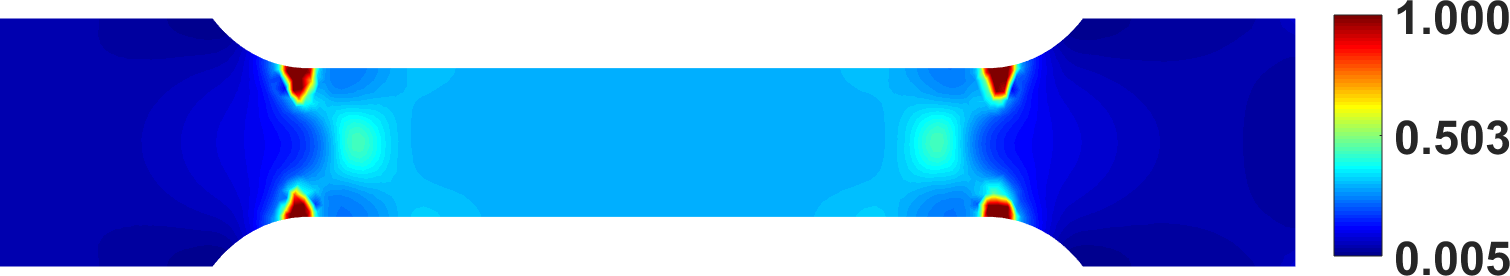}}
		\caption{Damage phase-field for each representative failure case. By changing the parameters $\gamma$, $g_c$, and $c$, we observe different failure types (distinct crack positions and paths), as well as varying dynamics.}
		\label{fig:cases_field}
	\end{figure}

\begin{figure}[t]
		\centering
		\subfloat[Case 1.]{\includegraphics[width=0.33\textwidth]{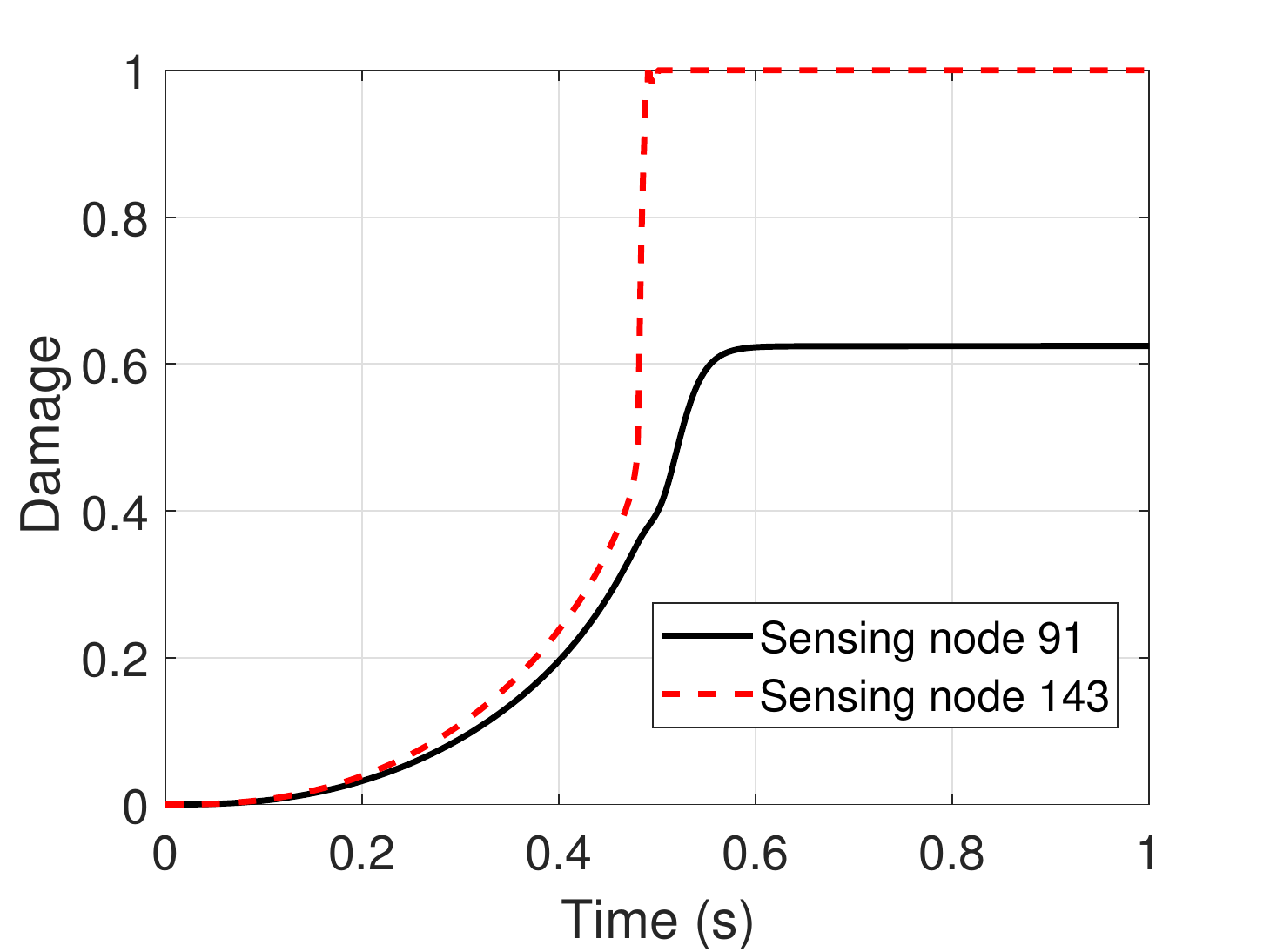}}
		\subfloat[Case 2.]{\includegraphics[width=0.33\textwidth]{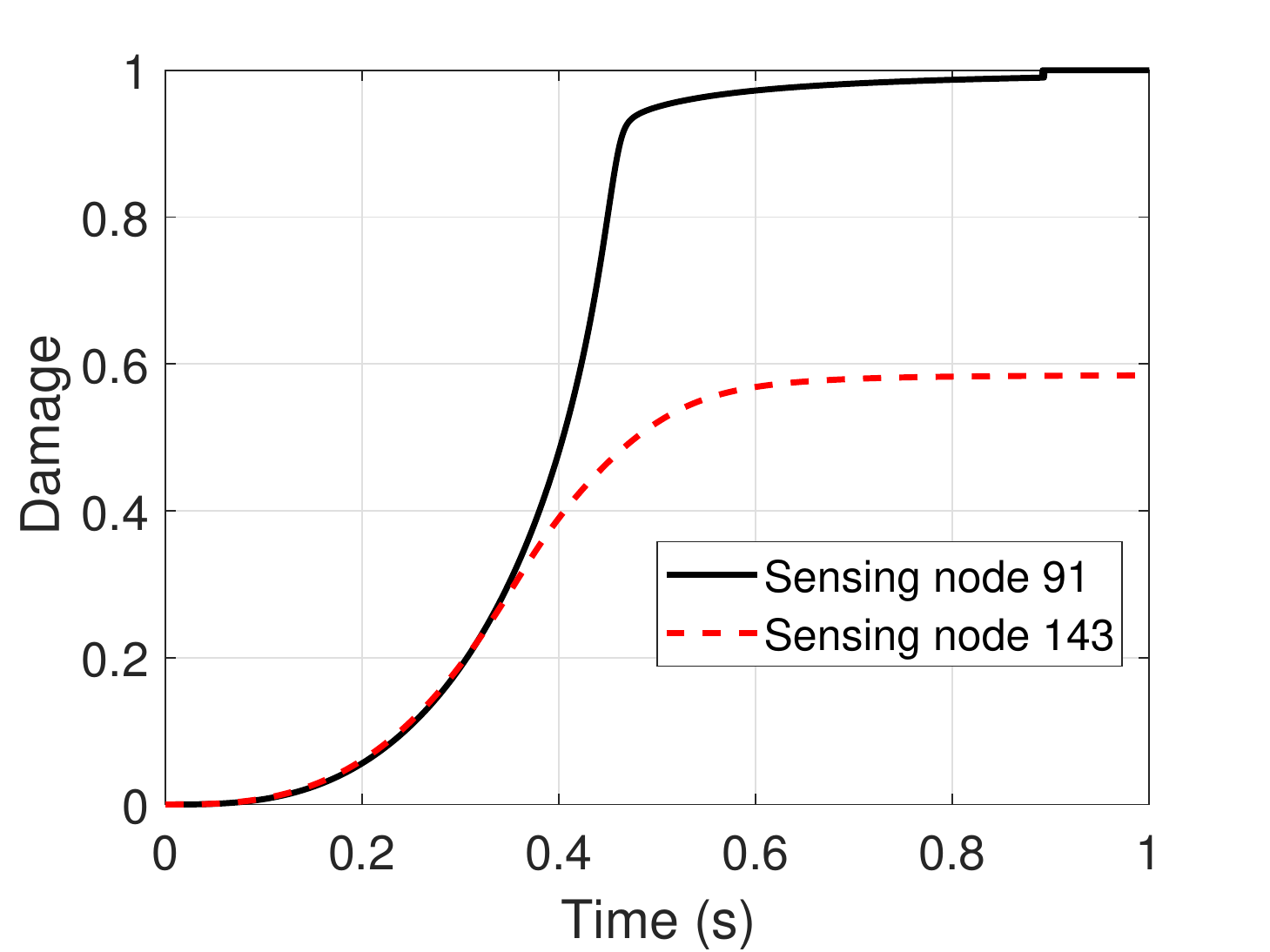}}
		\subfloat[Case 3.]{\includegraphics[width=0.33\textwidth]{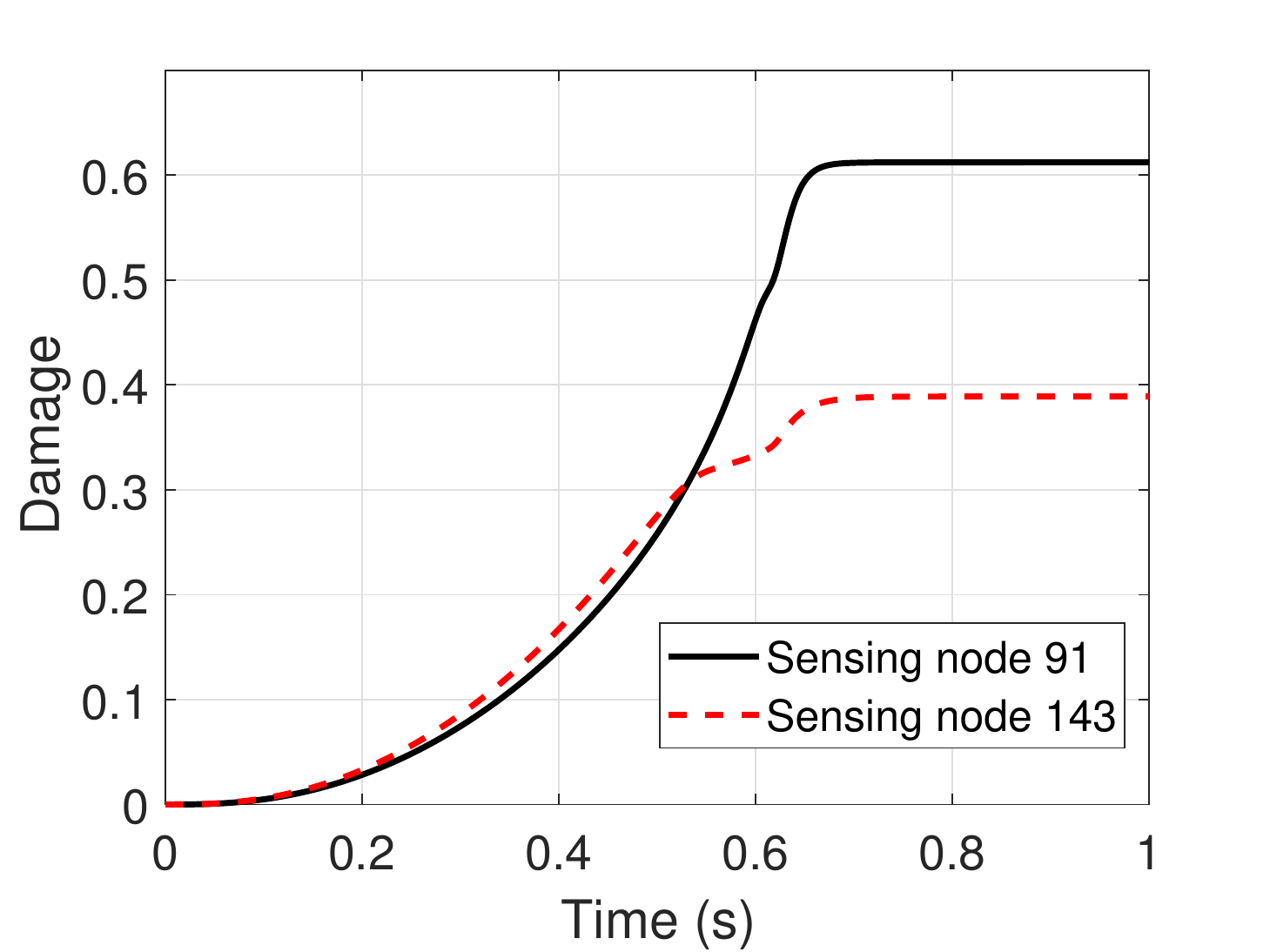}}
		\caption{Damage phase-field time-series data for three cases, showing the different evolution of $\varphi$ depending on the virtual sensor node position.}
		\label{fig:ts}
	\end{figure}

	\Cref{tab:cases} presents the parameters used to construct each representative failure case or type, for which we plot the damage phase-field at failure time in \Cref{fig:cases_field}. We also observe three different failure types (i.e., around the fillets, at the middle of the specimen, and in an intermediate region between those), where the effect of changing parameters is clearly noticeable. We observe the different damage evolution in the highlighted sensor nodes shown in \Cref{fig:geo_mesh} from their time-series data. We plot time-series data from sensor nodes 91 (at the middle of the specimen) and 143 (at one of the fillets), for cases 1, 2 and 3, where we observe the different evolution profiles based on each failure type (see \Cref{fig:ts}).
	
	\begin{table}[t]
		\centering\begin{threeparttable}
		\caption{Parameters used in the representative cases.}
		\label{tab:cases}
		\centering\begin{tabular}{|p{2cm}|p{3cm}|p{3cm}|p{3cm}|}\hline
			 Case & $\gamma\ (m)$ & $g_c\ (N/m)$ & $c\ (\frac{m}{Ns})$ \vspace{0.1cm}\\ \hline 
			1        & $3.00\times 10^{-4}$ & $2700$  & $2.00\times 10^{-6}$\\ \hline
			2        & $2.00\times 10^{-3}$ & $2700$  & $2.00\times 10^{-6 }$ \\ \hline
			3        & $5.00\times 10^{-4}$ & $5400$  & $2.00\times 10^{-6}$\\ \hline
			4        & $5.00\times 10^{-4}$ & $10800$ & $2.00\times 10^{-6}$ \\ \hline
			5       & $2.00\times 10^{-3}$ & $5400$  & $1.00\times 10^{-6}$ \\ \hline
			6       & $2.50\times 10^{-4}$ & $5400$  & $1.00\times 10^{-6}$\\ \hline
		\end{tabular}
	\end{threeparttable}
	\end{table}
	
	From the damage phase-field time-series data $\varphi$ at sensing nodes, we then form a feature vector of a pattern as the degradation function $g(\varphi) = (1-\varphi)^2$. Thus, patterns are generated using $g(\varphi)$, extracted at sensing nodes for a given time step. Accordingly, a label is assigned to identified patterns at each time step. 
	
	\begin{remark}
	Damage $\varphi$ is a proper measure of material failure. However, by defining the feature vector based on the degradation function $g(\varphi)$ instead, we directly measure the material softening, since $g(\varphi)$ is the field variable that degrades the constitutive model, thus reducing the component's load bearing capabilities. 
	\end{remark}
	
	\subsection{Label Definitions}
	\label{subsec:data_processing}
	
	In the domain of ML, label is defined as the output of the classification algorithm. We outline different criteria used to generate the labels for the supervised ML algorithms. In the context of failure analysis, labels should reflect the material's capacity to withstand loads, so a first rational choice is to define labels based on load-displacement curve. Besides, we generate labels based on a damage threshold concept with degradation function $g(\varphi)$. Based on noted criteria, each pattern is given a label corresponding to one of multiple classes, namely, no failure (class 1), onset of failure (class 2), and failure (class 3).
	
	\subsubsection{Label definition according to load-displacement curve}
	
	We start by defining the labels in a binary fashion, in order to observe the damage phase-field corresponding to a specific label transition. At each time step, a pattern is assigned $0$ if there is no failure, and $1$ if the specimen has fractured, such that we have a label vector $L = \{0\,\, 0\,\, \dots\,\, 0\,\, 1\,\, \dots\,\,  1\,\, 1\}^T$. We assign the labels based on load-displacement curve of the tensile test:
    
    \begin{itemize}
    	\item \textbf{Label Type 1}: labels are generated based on the maximum force at the load-displacement curve, which may induce to a failure criterion too soon.
    	\item \textbf{Label Type 2}: labels are generated according to the minimum derivative $df/du$, which could detect damage too late.
    	\item \textbf{Label Type 3}: labels are generated based on  $85\%$, $90\%$, and $95\%$ of maximum force at the load-displacement curve, which yields an intermediate behavior compared to Label types 1 and 2.
    \end{itemize} 

\begin{figure}[t]
	\begin{minipage}[b]{.49\textwidth} 
		\centering
		\subfloat[Load-displacement curve.]{\includegraphics[width=0.9\textwidth]{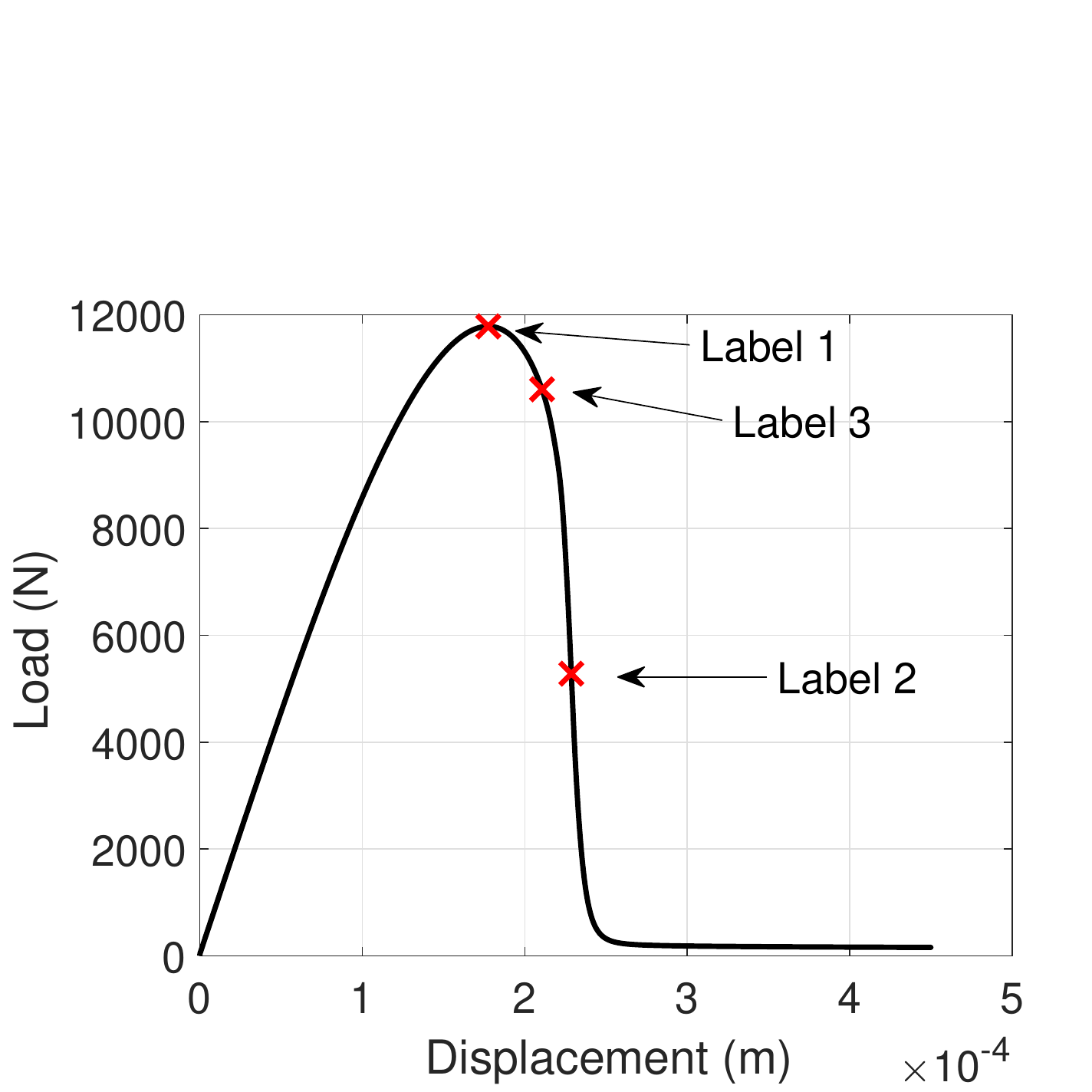}}
		\par\vspace{0pt}
    \end{minipage}
    \begin{minipage}[b]{.49\textwidth} 
		\centering
		\subfloat[Label 1, $t = 0.39\ s$.]{\includegraphics[width=0.9\textwidth]{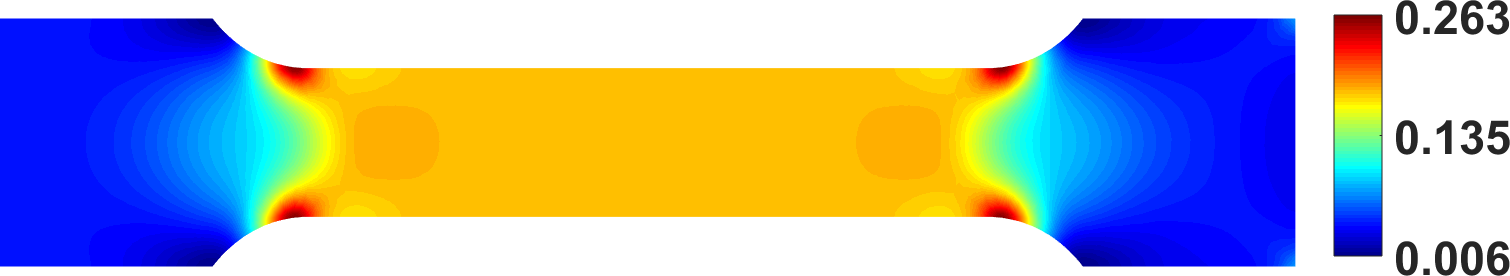}}\\
		\subfloat[Label 2, $t = 0.51\ s$..]{\includegraphics[width=0.9\textwidth]{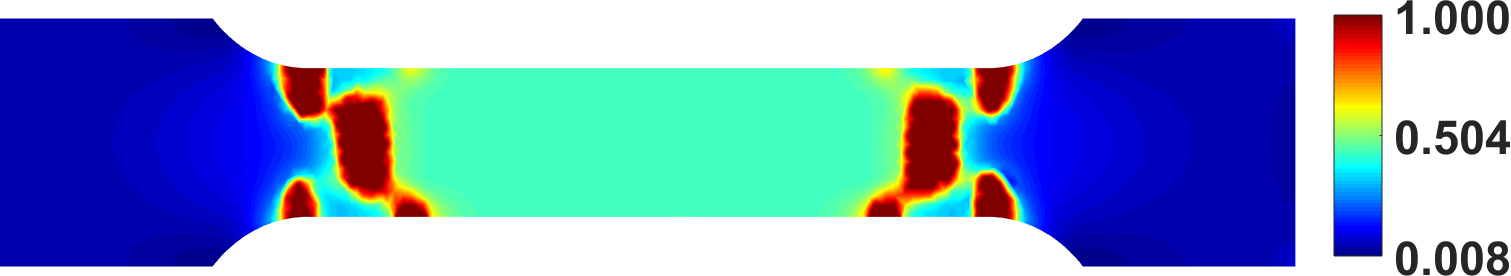}}\\
		\subfloat[Label 3, $t = 0.47\ s$..]{\includegraphics[width=0.9\textwidth]{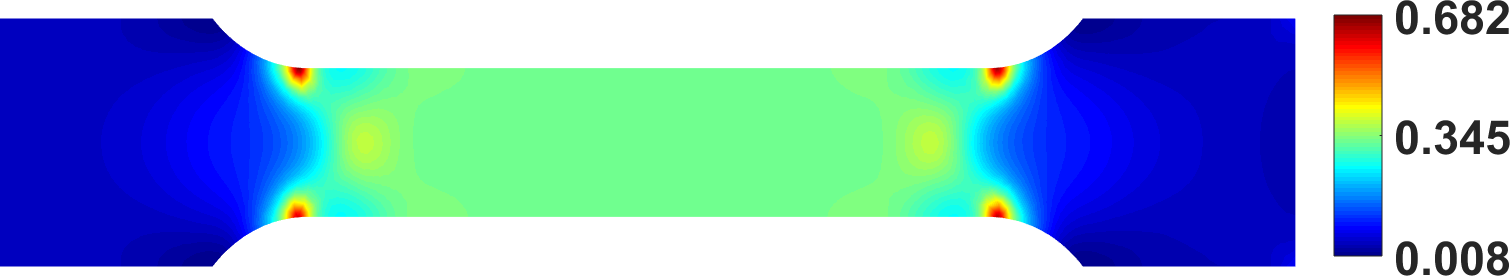}}\\
		\par\vspace{0pt}
	\end{minipage}
	    \caption{(Left) Load-displacement curve for case 1, where we identify the three points where the labels change from 0 to 1, according to different criteria. (Right) Respective damage phase-fields corresponding to the positions indicated in the curve. We note that Label type 3, based on a threshold of $90\%$ of maximum force, lies between the first two criteria. In Label 1, damage field is still too smooth, while in Label 2, failure is far too advanced.}
		\label{fig:labels}
	\end{figure}

    \Cref{fig:labels} illustrates the different points, where failure is defined according to the label types, with corresponding damage phase-fields. The label types 1 and 2 are too extreme and lead to early and late prediction of failure, respectively. To address this issue, we further improve binary label type 3 by including an intermediate state, the onset of failure, based on percentages of maximum load. Label type 3 then becomes a multiple label definition, stated here as:
    
    \begin{itemize}
    	\item \textbf{Multiple Label Type 3}: given the labels created based on $90\%$ and $95\%$ of maximum force at the load-displacement curve ($l_{90}$ and $l_{95}$), a pattern $x_i$ is assigned to class 1 (no failure) if label of the pattern based on $l_{90}$ and $l_{95}$ is zero, class 2 (onset of failure) when label of the pattern is zero based on $l_{90}$ and one based on $l_{95}$, and class 3 if label of the pattern based on $l_{90}$ and $l_{95}$ is one.  
    \end{itemize} 
	
	\subsubsection{Label definition according to damage threshold concept}
	We also propose a label definition based on a damage threshold of degradation function $g(\varphi)$, where three different thresholds (i.e., $R_1$=1, $R_2$=0.92, and $R_3$=0.85) are empirically selected based on the simulations. Accordingly, we generate labels by tracking $g(\varphi)$ on all sensing nodes, and following the rule:
	
	\begin{itemize}
    	\item \textbf{Multiple Label Type 4}: a given sensor node $S_i$ is shown with index $a$ when $R_1\geq g(\varphi) > R_2$, index $b$ if $R_2\geq g(\varphi) > R_3$, and index $c$ if $R_3\geq g(\varphi)$. Once the noted indices for all sensor nodes at a given time step (i.e., features of pattern $x_i$) are determined, sum of each index $a$ to $c$ is computed. Pattern $x_i$ is then classified as class 1, if summation of index $a$ is larger than that of indices $b$ and $c$, class 2 when summation of index $b$ is greater than summation of indices $a$ and $c$, etc. This label definition is motivated by the neighboring effect concept (i.e., group of sensor nodes), allowing to eliminate the effect of uncertainty and faulty sensor.
    \end{itemize}

	\section{ML Algorithmic Framework}
	\label{sec:framework}
\begin{figure}[t!]
        \centering
        \includegraphics[width=0.75\textwidth]{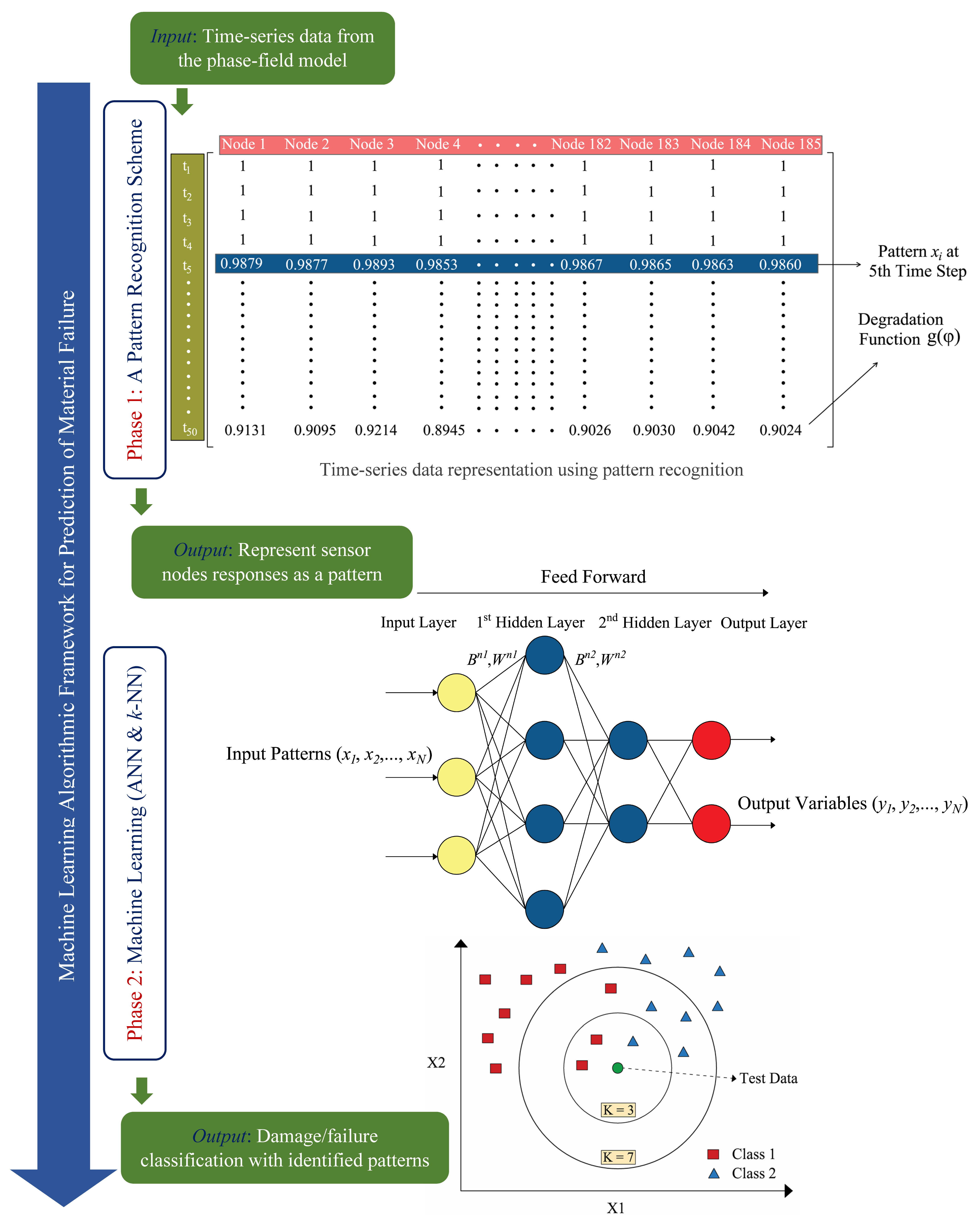}
        \caption{Schematic illustration of the proposed ML framework. A pattern recognition scheme is introduced to represent time-series data of damage degradation function $g(\varphi) = (1-\varphi)^2$ extracted at sensing nodes as a pattern. The $k$-NN and ANN algorithms are employed for failure classification using recognized patterns. In $k$-NN analysis the classification is performed by determining the $k$-nearest vote vector. An ANN provides a map between the inputs and outputs through determination of the weights using input and output patterns.}
        \label{fig:ML_framework}
    \end{figure}

We develop a supervised ML algorithmic framework for interpretation of time-series data generated from the phase-field model. The proposed ML framework presented in \Cref{fig:ML_framework} is based on the integration of a PR scheme and ML algorithms. According to PR scheme, sensor nodes responses (i.e., time-series data of degradation function $g(\varphi) = (1-\varphi)^2$) at each time step are represented as a pattern, along with corresponding label. The input to the learning framework is thus a matrix $M$ with dimension $m \times n$, where $m$ denotes the number of time steps and $n$ represents the number of sensor nodes. Consequently, each row of the matrix denotes a pattern, where the dimension of the PR problem is $n$ (i.e., a pattern with $n$ features). 

After generating the noted matrix $M$, ML algorithms, $k$-NN and ANN, are used for failure/damage classification with $m$ identified patterns. It should be noted that $k$-NN and ANN are used in this research due to their effective and reliable performance, while these algorithms are computationally efficient. The theoretical and mathematical details of $k$-NN and ANN can be found in the published literature \cite{keller1985fuzzy,hassoun1995fundamentals,zhang1998forecasting,weinberger2009distance}. The dataset for the $k$-NN and ANN analysis is divided into three subsets; namely, training, validation, and test. The training set is used to fit the ML classifiers, while the validation set is used to compute the optimal learning parameters. Performance of the ML classifiers with optimal parameters is then assessed on the test set. The performance of $k$-NN and ANN algorithms is measured using the detection performance rate defined in following equation:

\begin{equation}
\label{Eq:acc}
\text{Classification accuracy} = \frac{\text{Number of patterns correctly classified}}{\text{Total number of identified patterns}}.
\end{equation}

	Different size of data subsets are considered herein to evaluate the effect of such factor on the performance of the ML algorithms. Accordingly, five different combinations listed below are defined, where the accuracy of $k$-NN and ANN is determined based on each combination.

	\begin{itemize}
    \item Comb 1: training \& validation 65\% and test 35\%.
    \item Comb 2: training \& validation 70\% and test 30\%.
    \item Comb 3: training \& validation 75\% and test 25\%.
    \item Comb 4: training \& validation 80\% and test 20\%.
    \item Comb 5: training \& validation 85\% and test 15\%.
    \end{itemize}

	\section{Results and Discussion}
	\label{sec:results}
	The performance of the developed ML framework in terms of predicting the presence and location/pattern of failure is evaluated with time-series data of degradation function $g(\varphi) = (1-\varphi)^2$ generated from the phase-field model. To this aim, the framework is initially trained and tested using each one of six representative failure cases (see \Cref{subsec:time_series}), where the presence of failure is detected for each case, along with corresponding accuracy. In the next analysis phase, to detect the location of failure, ML algorithms (i.e., $k$-NN and ANN) are trained using data from all six failure cases, and the classification accuracy is determined on test data, leading to identification of the pattern of failure. The following subsections present the classification results of the ML framework employing multiple labels generated according to multiple label types 3 and 4 (see \Cref{subsec:data_processing}).

	\subsection{Results with $k$-NN}
	\label{subsec:results with KNN}
	\subsubsection{Detection of the Presence of Failure}
\begin{figure}[t!]
    \centering
    \subfloat[]{\includegraphics[width=0.45\textwidth]{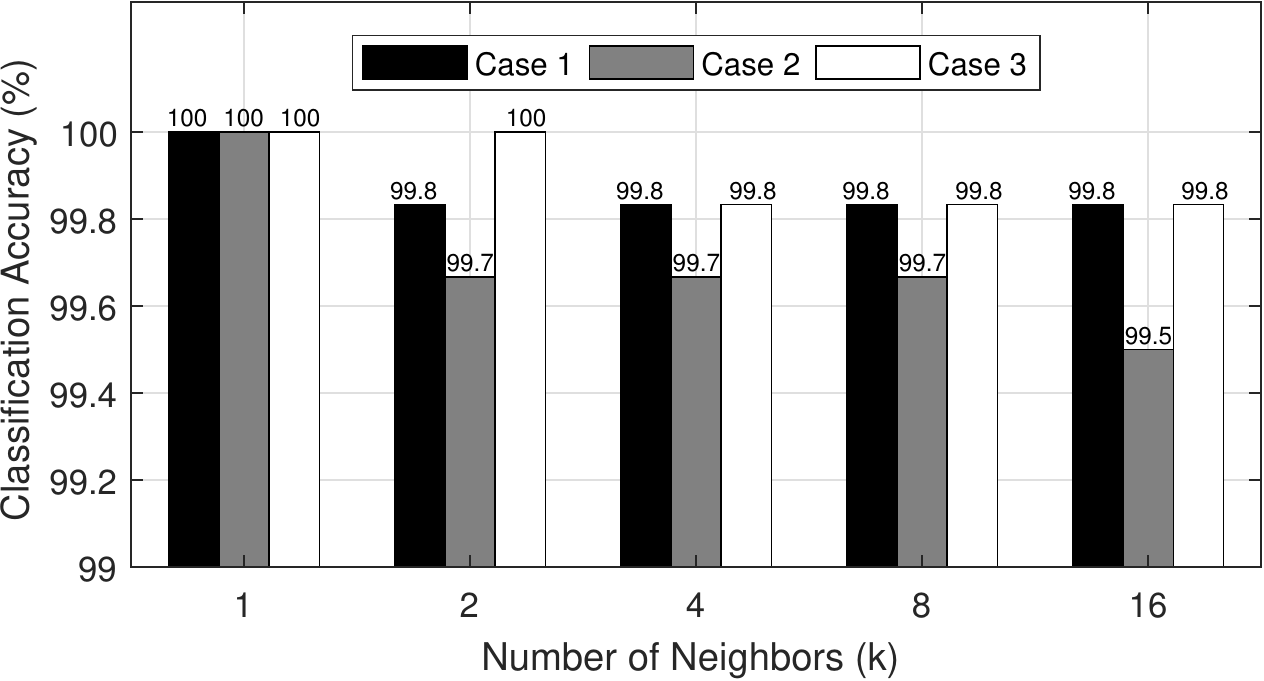}} \qquad
    \subfloat[]{\includegraphics[width=0.45\textwidth]{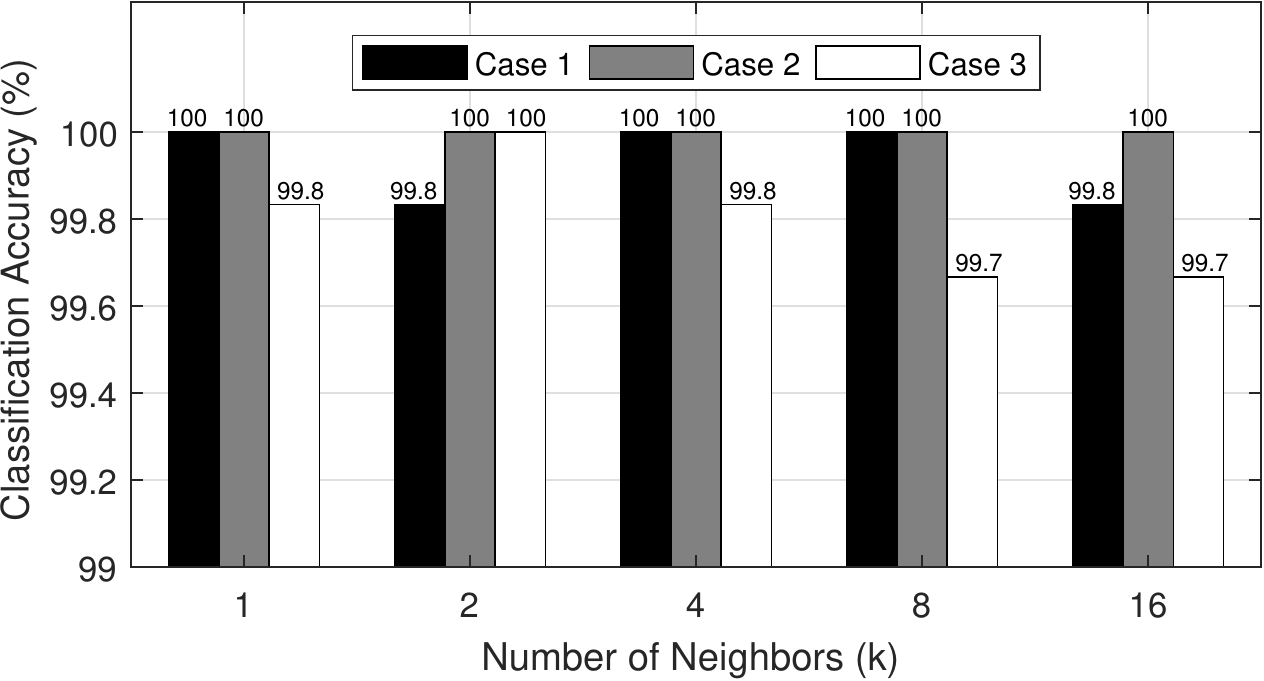}}
    \caption{$K$-NN classification accuracy with different number of $k$: (a) Accuracy based on multiple label Type 3, (b) Accuracy based on multiple label Type 4.}
    \label{fig:KNN_diff_k}
    \end{figure}

\begin{figure}[t!]
\centering
\includegraphics[width=0.5\textwidth]{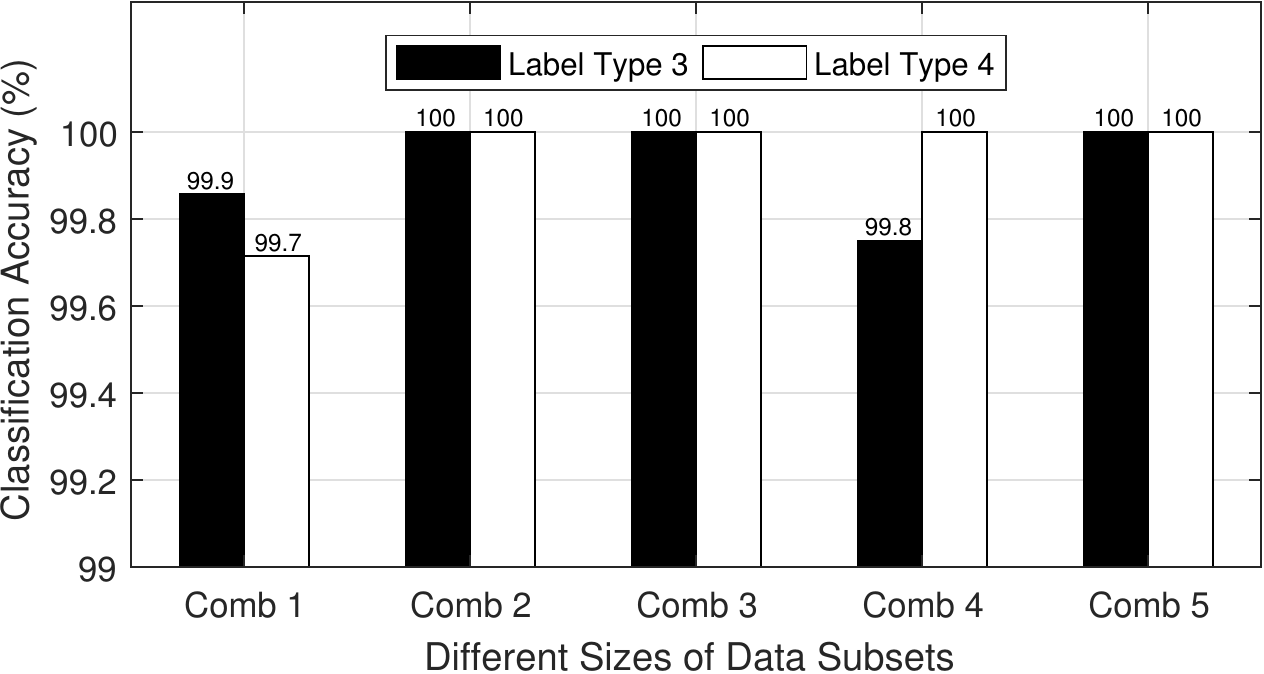}
\caption{$K$-NN classification results for failure case 3 with different size of data subsets and multiple label Types 3 \& 4.}
\label{fig:KNN_comb_results}
\end{figure}

	As noted in \Cref{sec:framework}, the proposed ML framework is trained and tested using different size of data subsets. For the $k$-NN analysis, $k$-fold cross validation with $k=10$ is used. Patterns identified with the PR scheme, along with corresponding labels, are used as input to the algorithmic framework in order to predict the presence of failure. To further explore the performance of the $k$-NN algorithm, the optimal number of $k$ needs to be determined. In this context, cases 1 to 3 representing different failure types/locations (see \Cref{fig:cases_field}) are considered, based on which the performance of the algorithm is evaluated with varying $k$. The $k$-NN classification results based on multiple label types 3 and 4 are shown in \Cref{fig:KNN_diff_k}. As can be seen, by increasing the number of neighbors ($k$), accuracy decreases. We choose $k = 2$ for subsequent analyses, and we will later check that this is the optimal number of neighbors in detection of failure location. Furthermore, the optimal distance is found to be ``Cosine", which results in better accuracy compared to other distance functions. Classification results with different size of data subsets are presented in \Cref{fig:KNN_comb_results}  for failure case 3, from which it can be observed that the highest accuracy is achieved based on combinations 2, 3 and 6, so we choose Comb 2, i.e., training \& validation 70\% , and test 30\% for all further results.
	
\begin{figure}[t!]
\centering
\includegraphics[width=0.52\textwidth]{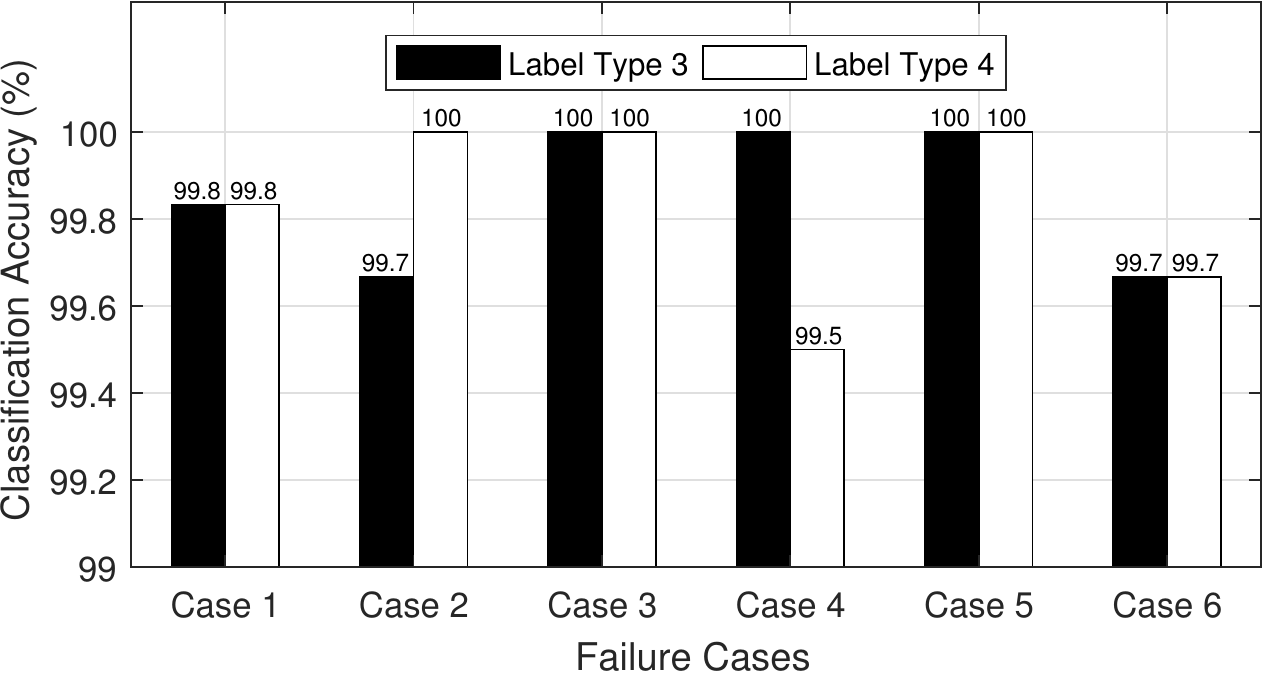}
\caption{$K$-NN classification results for different failure cases based on label Types 3 \& 4.}
\label{fig:KNN_diff_cases}
\end{figure}

\begin{figure}[t!]
		\centering
		\subfloat[]{\includegraphics[width=0.45\textwidth]{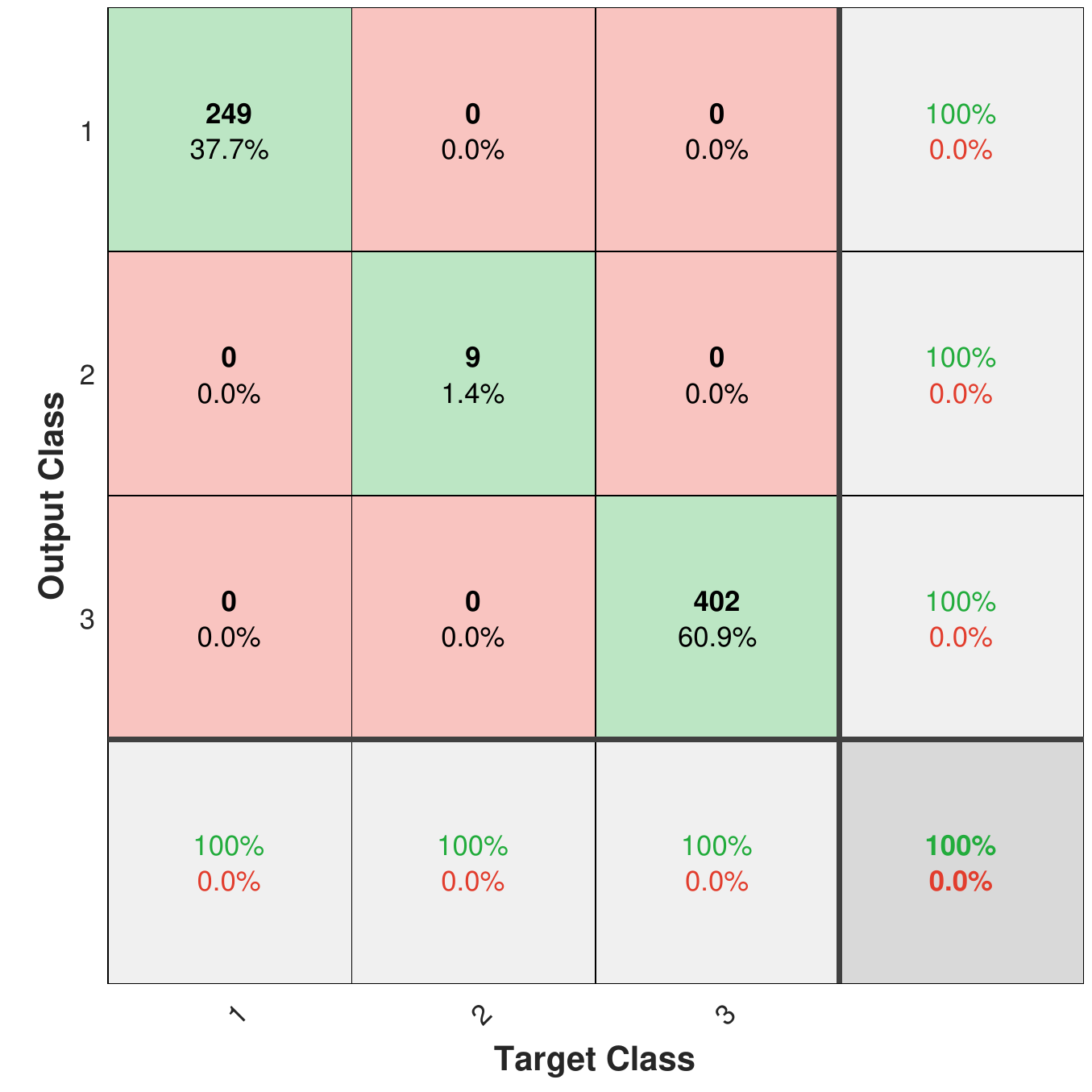}}
		\subfloat[]{\includegraphics[width=0.45\textwidth]{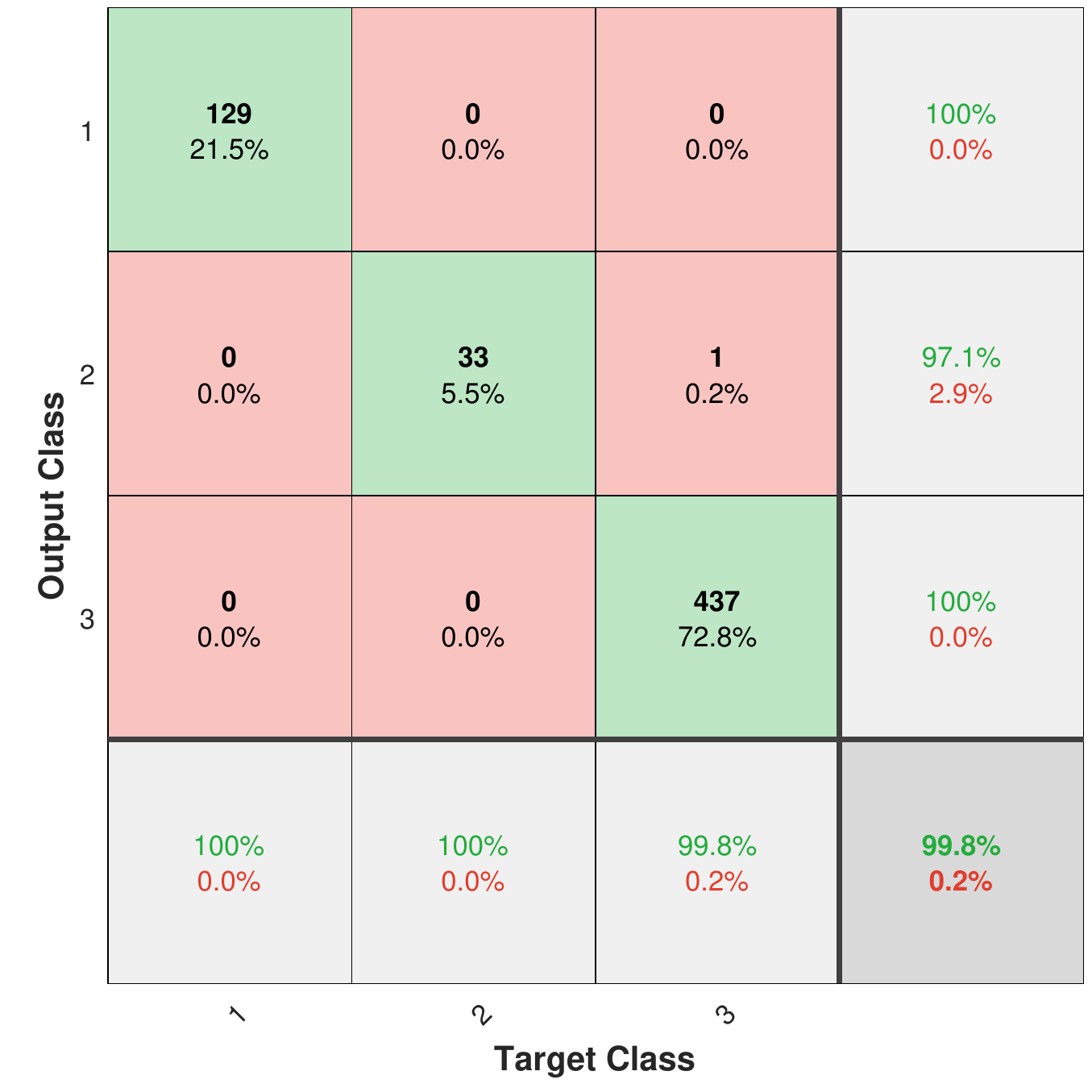}}
		\caption{Confusion matrix on test data with $k$-NN: (a) Case 5 and multiple label Type 3, (b) Case 1 and multiple label Type 4.}
		\label{fig:confmat_knn}
	\end{figure}

	The performance of the proposed ML framework employing $k$-NN with multiple labels is assessed for all failure cases (i.e., cases 1 to 6), where the classification accuracy on test data is reported for each case (see \Cref{fig:KNN_diff_cases}). Clearly, the performance of the framework is acceptable such that the highest accuracy based on multiple label types 3 and 4 is 100\% for failure cases 5 and 3, respectively. To better visualize the classification results, a confusion matrix containing information about actual and predicted classification is determined. Each column of the confusion matrix represents the patterns in a predicted class, whereas each row denotes the patterns in an actual class. The confusion matrix containing detailed classification results for cases 5 and 1 are depicted in \Cref{fig:confmat_knn}. As can be seen, the $k$-NN method performs well on all classes, including class 2 denoting onset of failure, which is of primary interest for early detection of failure in real-world applications.
	
	\subsubsection{Detection of the Location/Pattern of Failure}
\begin{figure}[t!]
	  	\centering
	  	\includegraphics[width=0.52\textwidth]{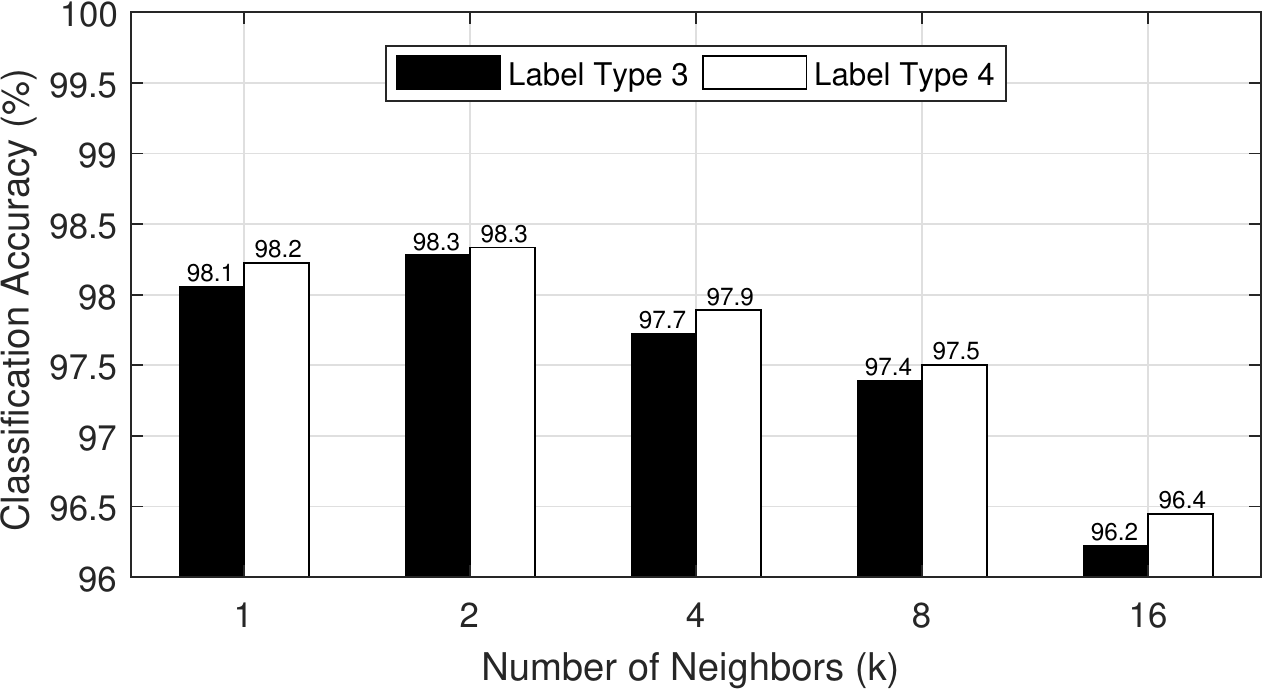}
	  	\caption{$K$-NN classification accuracy with different number of $k$ based on multiple label Types 3, and 4.}
	  	\label{fig:KNN_diff_k_pat}
	  \end{figure}

	  An attempt is made to detect the location of failure, enabling the framework to predict the pattern of failure. In regard to this, three different failure cases 1 to 3, representing different failure types, are considered for the analysis. Accordingly, nine classes/labels are defined, as shown in \Cref{tab:failure_labels}, using multiple label type 4 (multiple labels based on damage threshold concept). We also study the effect of different number of neighbors $k$ in this context. We present the accuracy results in \Cref{fig:KNN_diff_k_pat}, in which we observe that $k=2$ is indeed the optimal number of neighbors, corroborating the choice made in the previous section.

\begin{figure}[t!]
\centering
\includegraphics[width=0.6\textwidth]{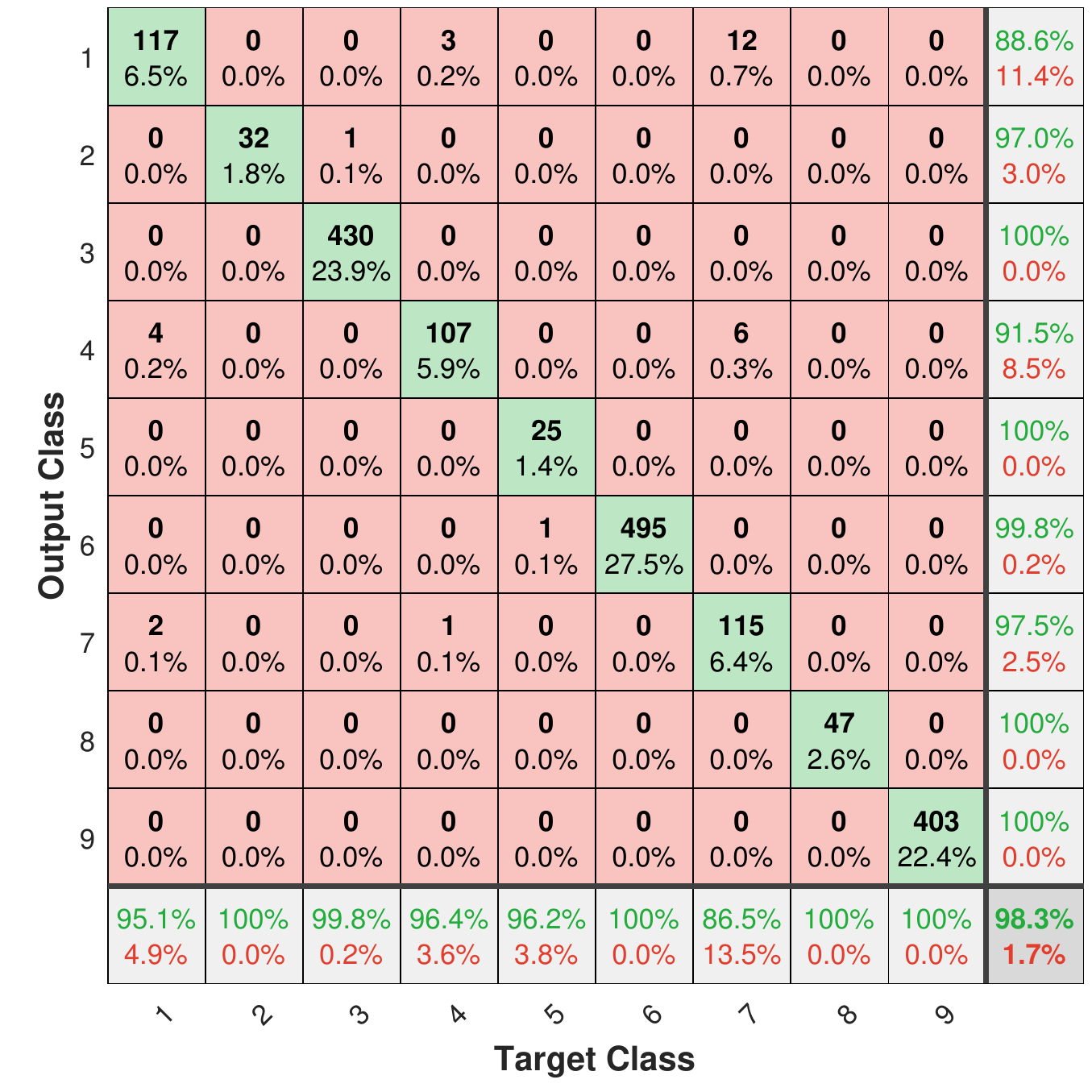}
\caption{Confusion matrix with $k$-NN classification results for detection of location/pattern of failure based on multiple label Type 4.}
\label{fig:confmat_KNN_failure_location}
\end{figure}

	  The confusion matrix showing the classification results is presented in \Cref{fig:confmat_KNN_failure_location}. As can be observed, the total accuracy is reported as 98.3\% using multiple labels (classes) shown in \Cref{tab:failure_labels} ($k$-NN detects the location of failure with 98.3\% accuracy). Results indicate the overall efficient performance of $k$-NN to detect the onset of failulre (classes 2, 5 and 8) and failure (classes 3, 6 and 9). The algorithm misclassifications are more concentrated in classes denoting no failure (classes 1, 4 and 7), where the method incorrectly identifies the location of failure in a few data points, because in early stages of the simulation the damage field is similar among the different cases. This is not an issue, since the critical part is the onse of failure. Moreover, the lowest classification accuracy is 86.5\% for class 7 (see \Cref{fig:confmat_KNN_failure_location}), which is an acceptable performance for a classification algorithm.
	
\begin{table}[t!]
\centering\begin{threeparttable}
\caption{Illustration of label/class definition for detection of location/pattern of failure.}
\label{tab:failure_labels}
\centering\begin{tabular}{|p{6cm}|p{1.2cm}|p{2.3cm}|p{1.2cm}|} \hline
Label/Class & Failure Case & Label based on Label Type 4 & New Label/ Class \\
\hline
& 1 & 0 & 1 \\ 
& 1 & 0.5 & 2 \\ 
& 1 & 1 & 3 \\ 
& 2 & 0 & 4 \\ 
Label of a Pattern at a Given Time Step & 2 & 0.5 & 5 \\ 
& 2 & 1 & 6 \\ 
& 3 & 0 & 7 \\ 
& 3 & 0.5 & 8 \\ 
& 3 & 1 & 9 \\ 
\hline
\end{tabular}
\end{threeparttable}
\end{table}
\makeatletter
\setlength{\@fptop}{0pt}
\makeatother

	\subsection{Results with ANN}
	\label{subsec:results with ANN}
	\subsubsection{Detection of the Presence of Failure}
\begin{figure}[t!]
\centering
\includegraphics[width=0.52\textwidth]{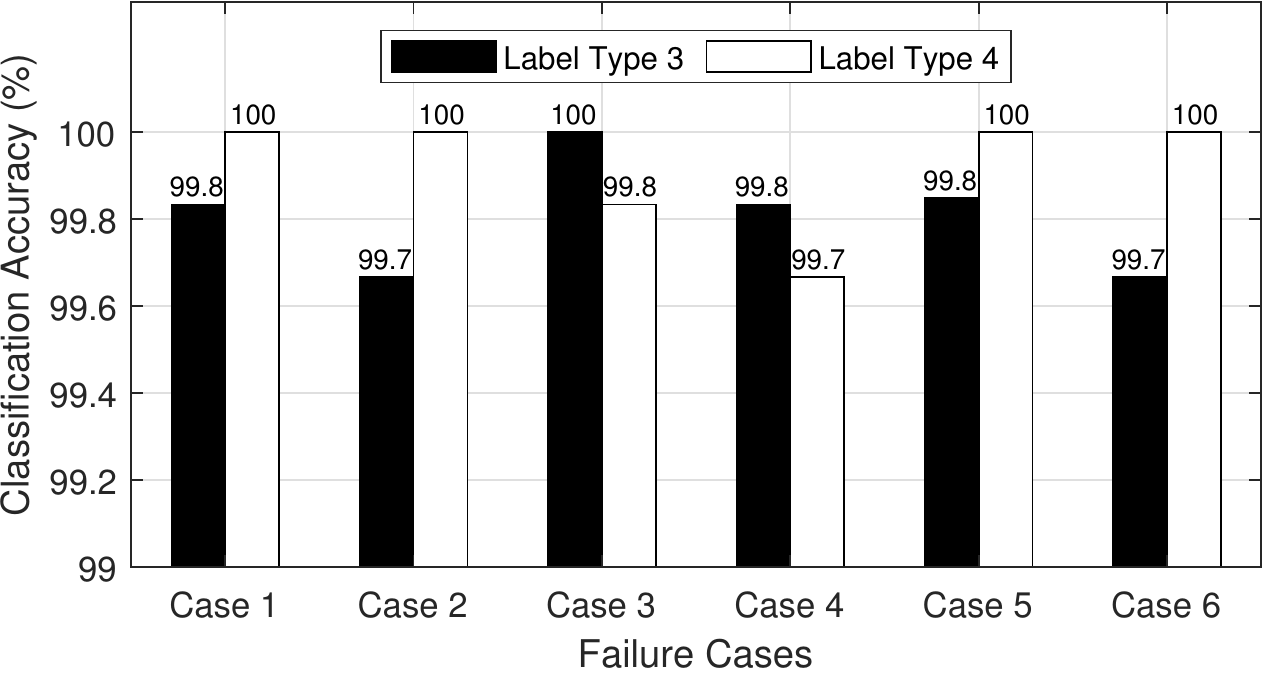}
\caption{ANN classification results for different failure cases based on multiple label Types 3 \& 4.}
\label{fig:ANN_diff_cases}
\end{figure}

\begin{figure}[t!]
		\centering
		\subfloat[]{\includegraphics[width=0.45\textwidth]{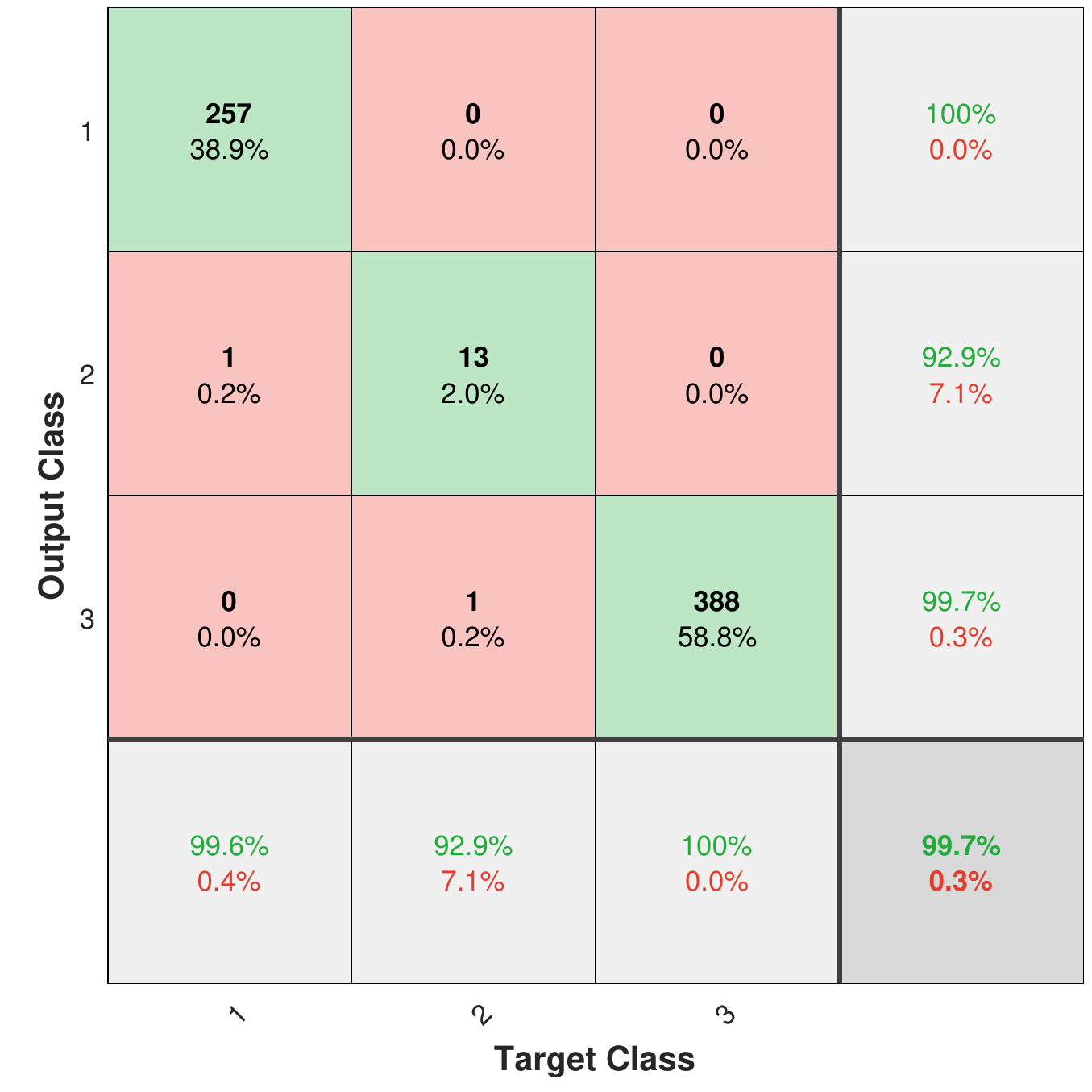}}
		\subfloat[]{\includegraphics[width=0.45\textwidth]{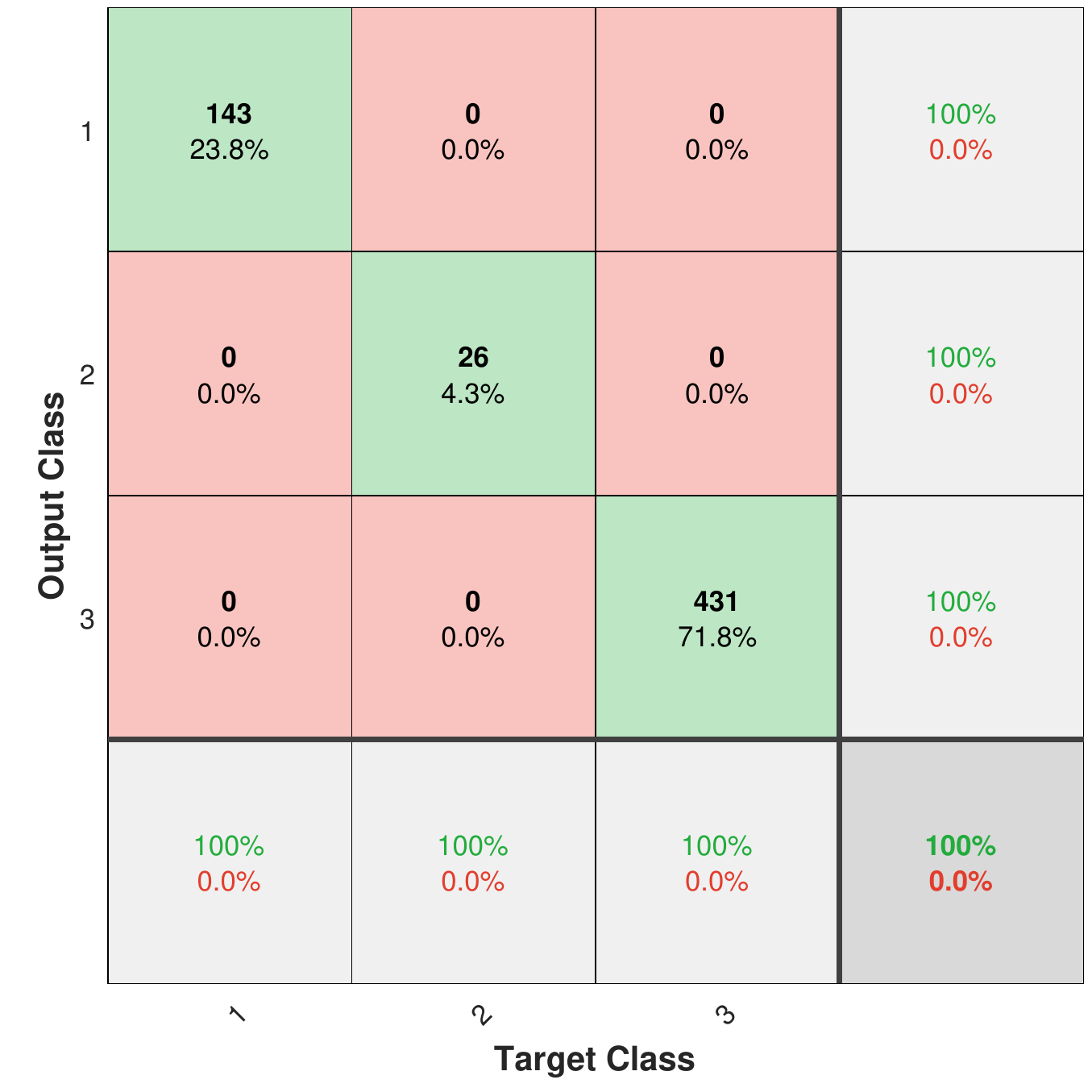}}
		\caption{Confusion matrix on test data with ANN: (a) Case 5 and multiple label Type 3, (b) Case 1 and multiple label Type 4.}
		\label{fig:confmat_ANN}
	\end{figure}

		The performance of the proposed ML framework employing ANN algorithm is evaluated in terms of detecting the presence of failure. On this basis, a two-layer (i.e., one hidden-layer with 5 neurons and output layer) feed-forward neural network with Sigmoid classifier/activation function is used. Classification results based on multiple label types 3 and 4 are presented in \Cref{fig:ANN_diff_cases}, from which it can be seen that ANN leads to comparable accuracy compared to $k$-NN (see \Cref{fig:KNN_diff_cases}). Detailed classification results with ANN is presented in a confusion matrix shown in \Cref{fig:confmat_ANN}. Results indicate that ANN effectively detects the presence of failure with total accuracy of 99.90\% and 100\% using multiple label types 3 and 4, respectively. 
		
	\subsubsection{Detection of the Location/Pattern of Failure }
\begin{figure}[t!]
\centering
\includegraphics[width=0.6\textwidth]{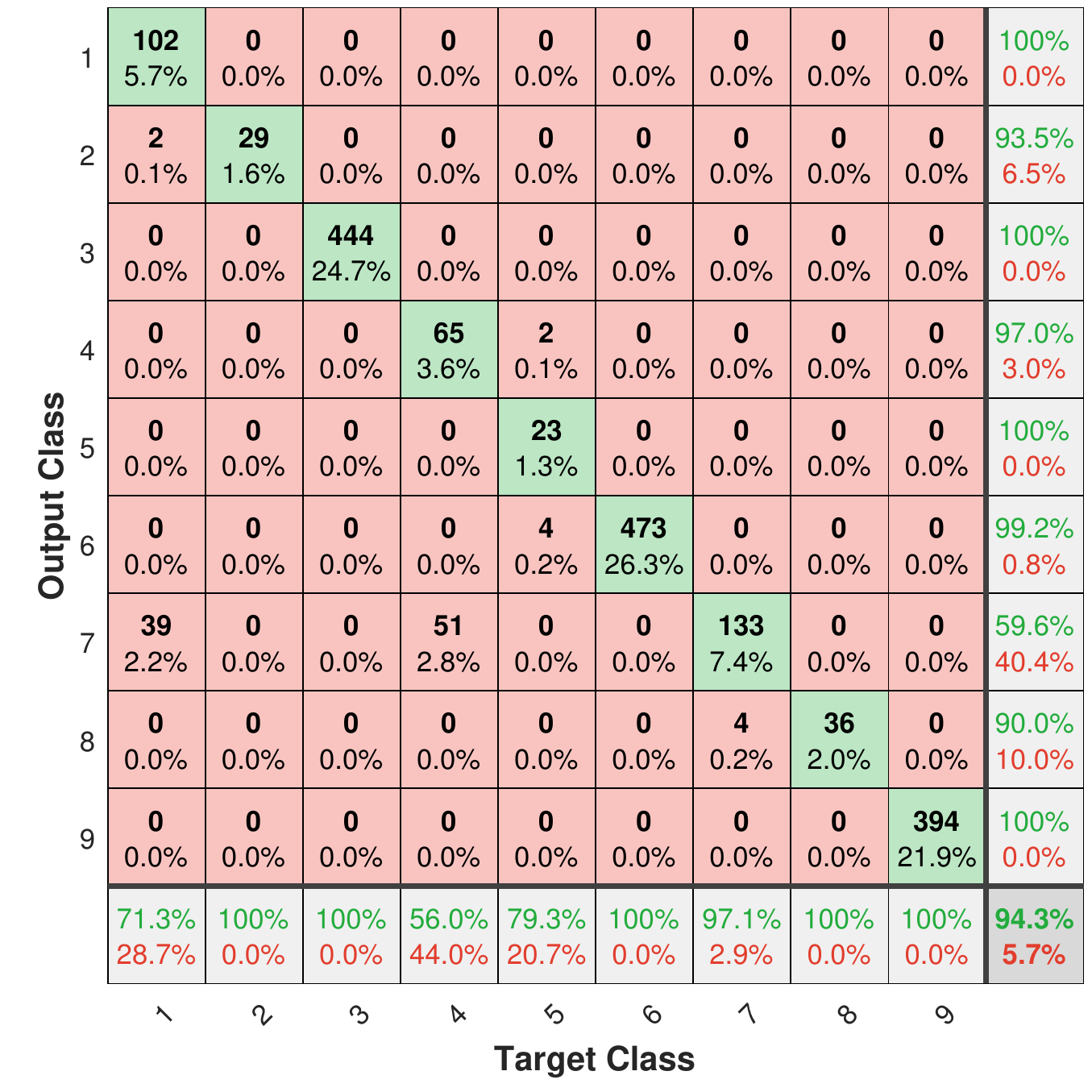}
\caption{Confusion matrix with ANN classification results for detection of location/pattern of failure based on multiple label Type 4.}
\label{fig:confmat_AA_failure_location}
\end{figure}

	Once the presence of failure is detected, ANN algorithm is employed to identify the location/pattern of failure. On this basis, multiple labels defined in \Cref{tab:failure_labels} are used for supervised classification with ANN algorithm. A two-layer (i.e., one hidden-layer with 5 neurons and output layer) feed-forward neural network with Sigmoid activation function is used as the ANN architecture to detect the location of failure. The confusion matrix with detailed classification results is presented in \Cref{fig:confmat_AA_failure_location}, from which it can observed that the location/pattern of failure can be successfully detected using ANN with high accuracy.  Similarly to $k$-NN, the majority of misclassifications in ANN belong to classes representing no failure (classes 1 and 4 in \Cref{fig:confmat_AA_failure_location}), due to the similarity of damage field prior to damage localization and crack initiation. In the other classes, ANN still performs successfully, with minimum accuracy of 79.3\% when detecting the onset of failure (class 5).

	\subsection{Uncertainty Quantification}

	The results presented in previous sections consisted in smooth, deterministic input data in a single run of the ML algorithms. In this section we propagate the uncertainty associated to data sampling and randomness associated with the algorithms, and we assess the robustness and accuracy of both methods to variability in data through the addition of Gaussian noise. This approach aims to verify the effectiveness of the framework to handle real-world data. 
		
	\subsubsection{Algorithmic randomness}
	
    We first study the propagation of uncertainties related to the algorithms, still using deterministic data. Such randomness appears in both $k-$NN and ANN due to the random division of time-series data into training, validation and test sets. We need to choose random division to avoid bias, specially in damage data that shows pronounced temporal evolution trends. Furthermore, ANN also presents another source of uncertainty, related to random initialization of weights and biases in each neuron.
	
   	We use Monte Carlo (MC) method to run multiple classification problems for each algorithm, and compute the expected total classification accuracy and standard deviation. We use multiple label Type 3, and run 1000 simulations for detection of failure presence in each case (cases 1 to 6), and run additional 1000 classifications for the detection of failure location, using classes from \Cref{tab:failure_labels}. We show the results in \Cref{tab:uq} . We observe that $k$-NN performs better than ANN in this setting, since ANN incorporates another level of uncertainty from the random guesses of neuron parameters. The randomness of data division does not affect the performance of neither method in the failure location problem. For failure location, ANN is less accurate, but still within accptable range.
   	
   	\begin{table}[]
   		\centering\begin{threeparttable}
   		\caption{Total classification accuracy mean and standard deviation from algorithmic randomness (\%).}
   		\label{tab:uq}
   		\centering\begin{tabular}{|l|c|c|c|c|}
   			\hline
   			\multirow{2}{*}{}              & \multicolumn{2}{c|}{$k$-NN} & \multicolumn{2}{c|}{ANN} \\ \cline{2-5} 
   			& Mean       & Std. Dev     & Mean      & Std. Dev     \\ \hline
   			Failure presence - Case 1      & 99.86      & 0.16         & 99.82     & 1.53         \\ \hline
   			Failure presence - Case 2      & 99.86      & 0.16         & 99.88     & 0.14         \\ \hline
   			Failure presence - Case 3      & 99.87      & 0.14         & 99.84     & 0.27         \\ \hline
   			Failure presence - Case 4      & 99.86      & 0.16         & 99.82     & 0.31         \\ \hline
   			Failure presence - Case 5      & 99.87      & 0.14         & 99.89     & 0.15         \\ \hline
   			Failure presence - Case 6      & 99.86      & 0.15         & 99.88     & 0.16         \\ \hline
   			Failure location - Cases 1/2/3 & 98.28      & 0.31         & 84.58     & 6.00         \\ \hline
   		\end{tabular}
   	\end{threeparttable}
   	\end{table}

    \subsubsection{Noisy data}
\begin{figure}[t!]
    	\centering
    	\subfloat[]{\includegraphics[width=0.45\textwidth]{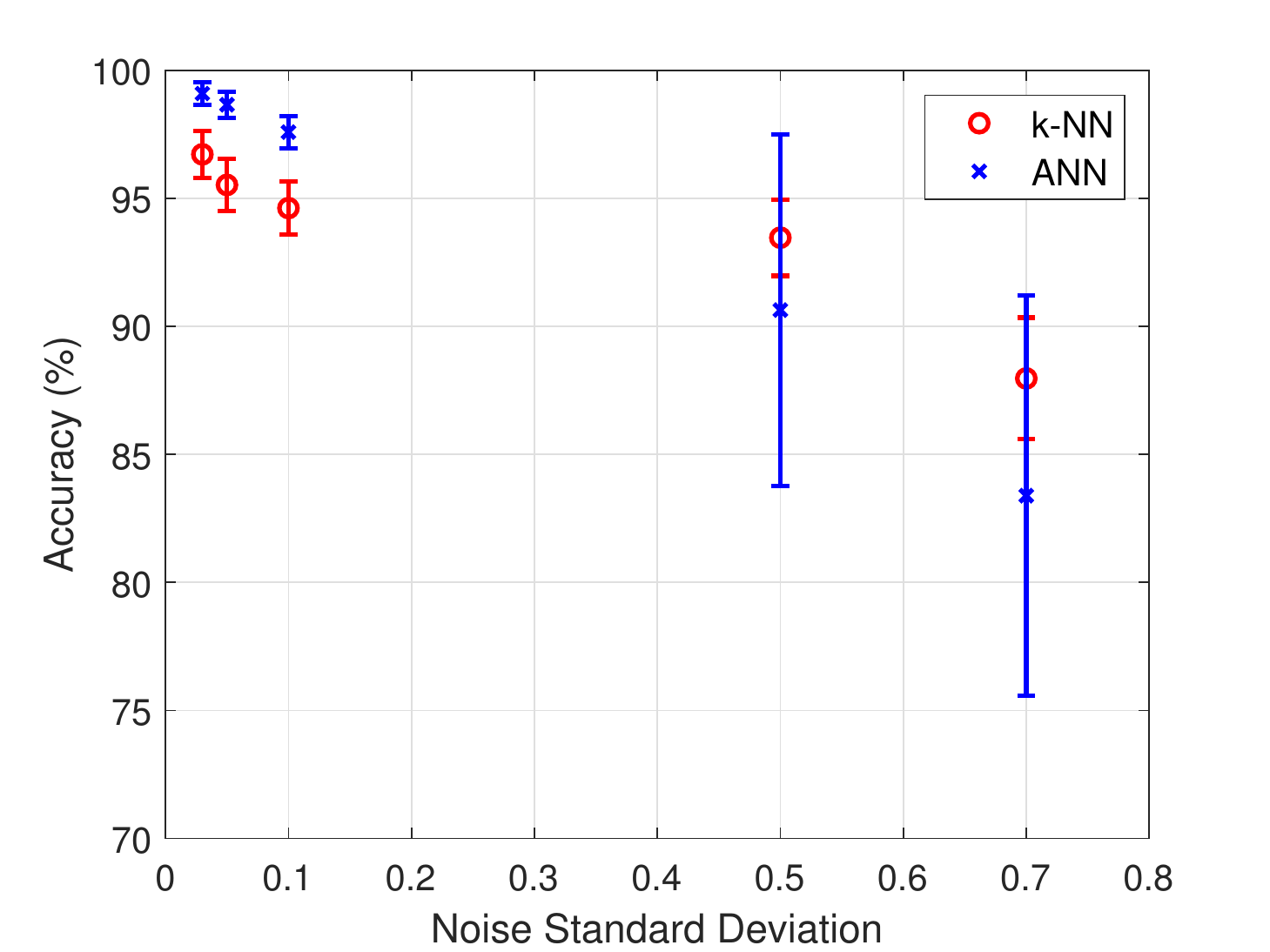}}
    	\subfloat[]{\includegraphics[width=0.45\textwidth]{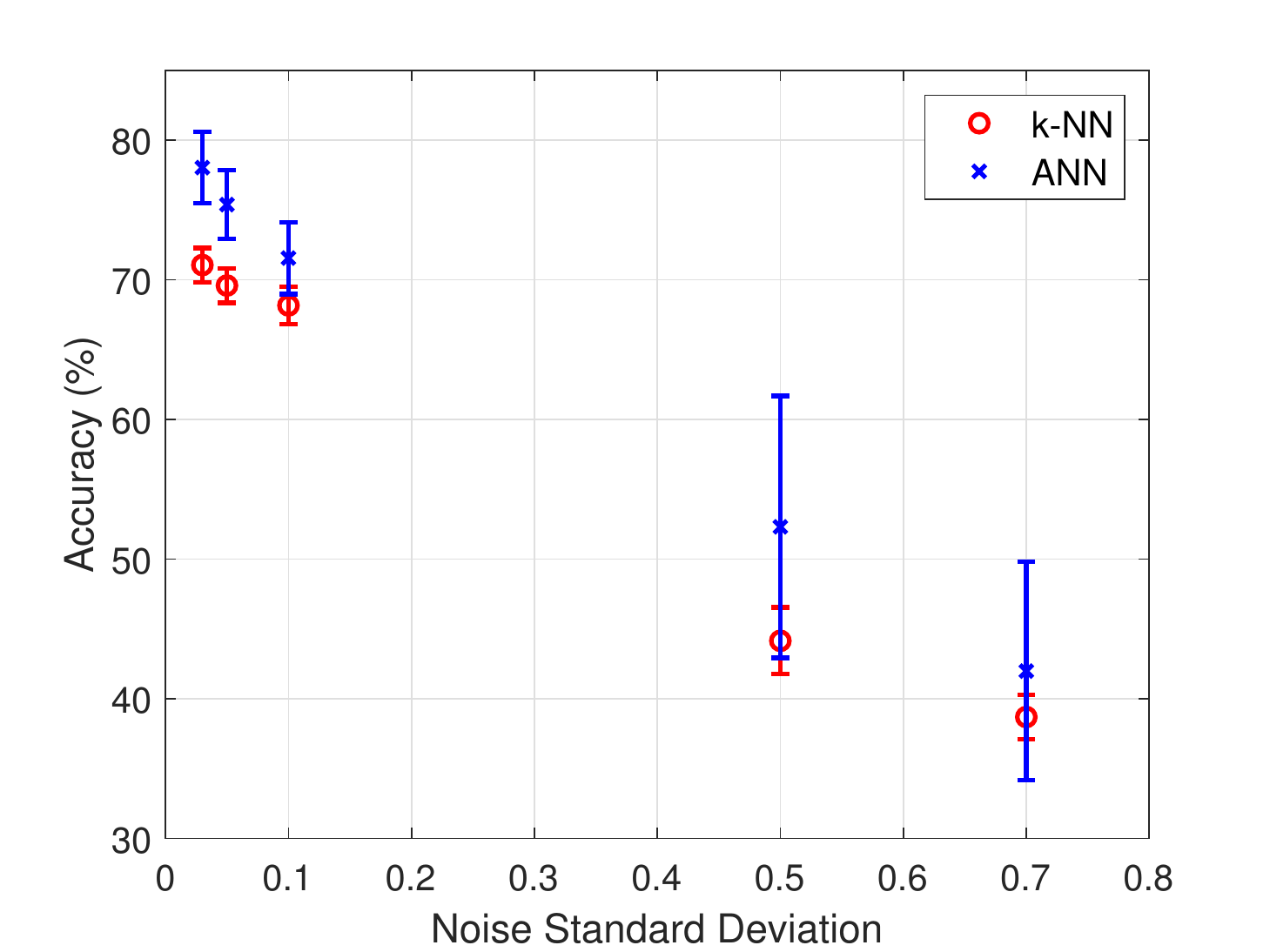}}
    	\caption{Mean total classification accuracy and standard deviation for (a) Detection of failure presence, case 3 and , (b) Detection of failure location.}
    	\label{fig:noise}
    \end{figure}

    To assess the performance of the proposed ML framework with noisy data, time-series data are corrupted by adding Gaussian distributed noise with different standard deviations to damage $\varphi$ data. We run 1000 MC simulations for  $k$-NN and ANN algorithms using multiple label Type 3, and compute the total accuracy expectation and standard deviation. We propagate uncertainty for failure presence detection in case 3, and for detection of failure location, and show the results in \Cref{fig:noise}.
    
   We observe that increasing noise levels decreases the mean accuracy, while enlarging the uncertainty range, for both methods. We see that even small intensity noise drops the accuracy to lower levels when compared to mean values based only on algorithmic randomness. In other words, input data dominates over algorithms' uncertainty. For failure presence detection in case 3, we see that ANN performs better than $k$-NN for low noise levels, while $k$-NN is more robust under higher noise magnitudes, resulting in higher mean accuracy and lower standard deviation. For case 3, the lowest accuracy was still above 75\%, showing robustness with 3 classes. Similarly, when we look at detection of failure location, ANN is superior $k$-NN for low noise levels, and shows higher mean accuracy even for high noise intensity, yet with more uncertainty. However, we cannot claim good performance of the algorithms with 9 classes under high noise levels, under this specific choice of algorithmic setup. This motivates a more systematic approach to uncertainty and sensitivity of ML algorithms under noisy phase-field data, and it shall be the focus of future studies.

	\section{Summary and Conclusions}
	\label{sec:conclusions}
	This paper presents a phase-field based machine learning (ML) framework developed to predict failure of brittle materials. Time-series data are generated according to nodal damage results from finite element simulations of a tensile test specimen. We assessed the performance of the proposed ML framework employing PR scheme and ML algorithms ($k$-NN and ANN) for different failure types, and with multiple labels generated based on load-displacement curve and damage threshold concept. We draw the following conclusions from the carried out study:
	
	    \begin{itemize}
    	\item Results indicate the acceptable performance of the proposed framework with multiple labels, in which a PR scheme is effectively used to represent time-series data of degradation function $g(\varphi) = (1-\varphi)^2$ as a pattern. This choice of time-series data is effective since it directly complies with the material softening behavior.
    	\item Both $k$-NN and ANN were efficient to predict the presence and location of failure. The majority of errors in detection of failure location were concentrated in classes representing no failure, due to smoothness and similarity of damage field early in the simulations.
    	\item Uncertainty related to input data noise dominates over algorithmic randomness uncertainty. The framework showed robustness to noise when detecting failure presence, and showed acceptable accuracy with low noise levels when predicting failure location. In general, with noisy data ANN outperforms $k$-NN.
        \end{itemize}
	
	Results of this study demonstrate the satisfactory performance of the developed algorithmic framework and the applicability of ML for failure prediction with damage phase-field time-series data. This study aims to promote the state-of-the-art in data-driven failure prediction indicating the practical application of ML for failure detection in brittle materials. Findings from this study are expected to advance the development of data-driven frameworks capable of establishing a direct relation between the classification accuracy and damage phase-field parameters. In other words, in such frameworks the output of ML framework can be further used as input to the damage phase-field model to identify the parameters leading to failure/damage. This will significantly result in enhancing the data-driven failure prediction framework's performance.
	
	It is acknowledged that the results presented in this paper are based on time-series data extracted at virtual sensing nodes across the test specimen. Yet, the proposed ML framework can be effectively applied to a system of real (i.e., non-virtual) sensing nodes. It is expected that the developed algorithmic framework will perform satisfactory with real sensor data. 
	
	It is worth pointing out that although the mathematical approach presented in this research is deterministic, there is an uncertainty associated with the proposed ML framework because of random nature of such intelligent algorithms. In addition, time-series data used in this study are smooth. Nonetheless, this issue cannot always be valid since near failure/damage regions time-series data can become noisy due to stress/strain concentration in such regions. Future works will focus on addressing the noted issues. Besides, the focus of future studies will be on the evaluation of the developed framework with the inclusion of plasticity/visco-elasto-plasticity \cite{suzuki2014transient,suzuki2016fractional,varghaei2019vibration,suzuki2019thermodynamically} in the damage and fatigue phase-field model, combined with efficient and stable long time integration schemes \cite{suzuki2018automated,zhou2019fast}.

	\bibliographystyle{siamplain}
	\bibliography{reference}
\end{document}